\documentclass[reprint,10pt,aps,prd,twocolumn,floats,floatfix,showpacs,superscriptaddress,nofootinbib]{revtex4}

\usepackage{amssymb,amsmath,amsfonts,amsthm,bm,graphicx}
\usepackage{multirow,txfonts}
\usepackage[usenames]{color}
\usepackage[dvipsnames]{xcolor}
\usepackage[utf8]{inputenc}
\usepackage[percent]{overpic}
\usepackage{aas_macros}
\usepackage[normalem]{ulem}
\usepackage{float}
\usepackage{lipsum}

\usepackage{subcaption}
\xdefinecolor{mylinkcolor}{rgb}{0,0,0.5}
\usepackage[
    bookmarksnumbered, bookmarksopen, bookmarksopenlevel=2,
    breaklinks=true,
    colorlinks=true, filecolor=mylinkcolor, citecolor=mylinkcolor,
    linkcolor=mylinkcolor, urlcolor=mylinkcolor, menucolor=mylinkcolor,
]{hyperref}
\usepackage{cleveref}

\allowdisplaybreaks

\begin{document}

\title{Dynamical transition to  spontaneous scalarization in neutron stars: The massive scalar field scenario}

\author{Juan Carlos Degollado}
\email{jcdegollado@icf.unam.mx}
\affiliation{Instituto de Ciencias F\'isicas, Universidad Nacional Aut\'onoma de M\'exico, Apartado Postal 48-3, 62251, Cuernavaca, Morelos, M\'exico}

\author{N\'{e}stor Ortiz}
\email{nestor.ortiz@nucleares.unam.mx}
\affiliation{Instituto de Ciencias Nucleares, Universidad Nacional Aut\'onoma de M\'exico, Circuito Exterior C.U., A.P. 70-543, M\'exico D.F. 04510, M\'exico.}

\author{Marcelo Salgado}
\email{marcelo@nucleares.unam.mx}
\affiliation{Instituto de Ciencias Nucleares, Universidad Nacional Aut\'onoma de M\'exico, Circuito Exterior C.U., A.P. 70-543, M\'exico D.F. 04510, M\'exico.}

\begin{abstract}
We analyze numerically the dynamical transition to spontaneous scalarization in neutron stars in the framework of a scalar-tensor theory of gravity where the scalar field is free but massive, and it is coupled nonminimally to gravity in the Jordan frame. We show that the quasistatic configuration of the star that settles after the transition can avoid the observational constraints imposed on the amount of scalarization by several observations in binary systems due to the presence of the mass term, which suppresses the range of the scalar field. We also study the impact of the scalar field mass on the gravitational mass of the star relative to the massless scenario.
\end{abstract}


\maketitle

\section{Introduction}
\label{sec:introduction}
Scalar-tensor theories of gravity (STT) are one of the simplest alternatives to Einstein's theory of general relativity (GR) (see Refs. \cite{Damour:1992we,Fujii2003} for a review). 
These theories introduce a new fundamental (real-valued) scalar field, $\phi$, apart from the metric tensor, $g_{ab}$, in the description of gravity. A prominent example of this kind of theories is the Jordan-Brans-Dicke theory \cite{Brans:1961sx,Brans:1962zz,Dicke62}, which has inspired a large amount of research during the past six decades or so, and provides a covariant mechanism to generate a variation of Newton's gravitational ``constant'' $G$ in terms of the scalar field $\phi$. This kind of variation, specially a cosmic temporal variation of $G_{\rm eff}$, was proposed during the 1990s to explain an apparent spatial periodicity in the distribution of galaxies around our own galaxy \cite{Broadhurst:1990be,Morikawa90}. As a bonus, this concocted model accounted for the dark sector of the universe (in the large scale), at least at the background level, prior to the discovery of the current accelerated expansion of the universe via the observation of supernovae SNIa, and prior to the cosmic microwave background (CMB) evidence on the presence of a cosmological constant \cite{Salgado96,Salgado97a,Salgado97b}. Later, STT were also analyzed more thoroughly in cosmological settings \cite{Boisseau:2000pr,Amendola2001,Riazuelo2002,Schimd2005}.

STT are usually formulated in two equivalent, but {\it philosophically} different fashions. One is the formulation in the so-called {\it Jordan frame} (JF), where the scalar field $\phi$ appears nonminimally coupled (NMC) to gravity, but with no direct coupling to the matter sector, and therefore, test particles follow geodesics of the so-called {\it physical metric} or JF metric, and the corresponding energy-momentum tensor (EMT) of matter $T_{ab}$ is conserved. The other formulation, the {\it Einstein-frame} (EF) or conformal frame representation of STT, is concerned with a scalar field $\tilde{\phi}$, related to $\phi$, which couples minimally to a conformal metric $\tilde{g}_{ab}$, but nonminimally to the corresponding EMT, $\tilde{T}_{ab}$ (see the Appendix \ref{app:einstein_frame}). Under the latter formulation, test particles do not (in general) follow geodesics associated with the conformal metric, and thus, its corresponding EMT is not conserved, in general.
Both representations have their own advantages, but the underlying physical effects (at least at the classical as opposed to the quantum level) are independent of the frame employed in the analysis \cite{Flanagan:2004bz}. In this paper we will focus mainly on the JF formulation, although, for practical reasons, we use the EF metric in some numerical calculations, as described in Secs.~\ref{sec:static_solutions} and \ref{sec:dynamics}.
An important feature of STT is that they have a well posed Cauchy (initial value) problem, notably, when formulated in the JF \cite{Salgado06,Salgado08}.

From the phenomenological point of view, STT faces different challenges as basically any alternative to GR. The stringent bounds imposed by several local observations (\textit{e.g.} solar system tests), as well as cosmological (\textit{e.g.} the CMB) and astrophysical observations (\textit{e.g.} binary systems, including binary pulsars) limit its possible deviations from GR due to the scalar field $\phi$ \cite{Berti:2015itd}.
Surprisingly, the theoretical discovery of the phenomenon of {\it spontaneous scalarization} (SC) in neutron stars under the framework of STT, by Damour and Esposito-Far\`ese \cite{Damour93} in the early 1990s, predicted deviations from GR in the strong-gravity regime while avoiding local and cosmic bounds. This phenomenon is analogue of {\it spontaneous magnetization} in ferromagnets at low temperatures \cite{Damour96}, in the sense that under certain conditions on the compactness of a neutron star, an energetically preferred configuration appears ``suddenly'' within the object, which is accompanied by the presence of a nontrivial scalar field cloud around the star while keeping the total baryon number fixed. This scalar field is endowed with an {\it order parameter} termed {\it scalar charge}, analogue to {\it magnetization} in ferromagnets.
The nontrivial scalar field does not require sources (\textit{i.e.} potentials) and vanishes asymptotically. This phenomenon is immune to solar system tests, and also to cosmological bounds, first because it does not appear around the sun due to its low compactness, and second, because it does not require a cosmological value, as the scalar field vanishes asymptotically.\footnote{The {\it induced scalarization}, as opposed to {\it spontateous scalarization}, corresponds to a phenomenon in compact objects where a background scalar field, \textit{i.e.} a nonzero asymptotic value for $\phi$, is present. In this case, one has to take into account the bounds imposed on local observations, like the bounds on the post-Newtonian parameters in the solar system due to a nonzero effective Brans-Dicke parameter $\omega_{BD}$, which has to be larger than $\sim 4\times 10^4$ in order to avoid the observational bounds (see Sec. \ref{sec:STT}). Nevertheless, in the spontaneous scalarization scenario, where $\phi\rightarrow 0$ asymptotically, $\omega_{BD}\rightarrow \infty$.}

The SC phenomenon is a nonperturbative effect \cite{Damour93} which in principle provides a new avenue for testing STT. In fact, STT also predicts a new polarization mode of gravitational waves (GWs), termed {\it breathing mode}, which leads to isotropic distortions of spacetime at the passage of GWs (even in the direction of propagation, like longitudinal waves do in matter) and which may be generated even in spherical symmetry. However, this does not occur in GR, where the predicted (transverse and traceless) GWs are sourced only by distributions of matter with quadrupole time-variations, and thus which deviate from spherical symmetry.
Moreover, GWs predicted by GR distort the spacetime only in the directions orthogonal to the passage of waves under the form of two independent modes ($++,\times \times)$. This feature has been corroborated systematically by the LIGO-VIRGO-KAGRA collaboration since 2014 \cite{LVK}, which has detected GW signals mainly sourced by the inspiriling and subsequent collision of binary black holes.
Regarding the simplest scenarios within GR, and even in STT, black holes (BHs) do not admit {\it hair}, notably in the form of a scalar field. This feature is consistent with current GW observations, which do not show evidence of a scalar field, whose presence might distort the predicted GW patterns. However, the story is different with neutron stars (NSs). In principle, the dynamical transition to SC could lead to the presence of scalar GWs that might be detected directly by future observatories sensitive to the breathing mode, or indirectly by showing distortions in the patters of the (usual) tensor modes of GWs \cite{Sotani05}. The first numerical analysis showing that such scalar GWs could be emitted during the transition to SC was performed by Novak \cite{Novak:1998rk}. Moreover, he showed that such waves can be further emitted during the gravitational collapse of a scalarized NS into a BH \cite{Novak98a}. Later, it was discovered that SC can also take place in boson stars \cite{Whinnett00,Alcubierre:2010ea}, and more recently, even in BHs in particular theoretical contexts, notably in scalar-Gauss-Bonnet gravity \cite{Silva:2017uqg,Doneva:2017bvd}. For a broad review on the phenomenon of SC in diverse compact objects and theories, see \cite{Doneva:2022ewd} and references therein.

Another, not less astonishing, consequence of the phenomenon of SC is that the maximum mass of NSs within STT can be larger than in GR \cite{Damour93,Salgado98}. This might be put forward to explain the existence of recently observed NSs with masses larger than two solar masses \cite{Antoniadis2013,Demorest2010,Cromartie2019,Romani2022}, without the need of exotic or very stiff equations of state. Despite these remarkable predictions, STT may seem to conflict with observations in binary pulsars, and more generally in binary systems \cite{Damour96,Freire2012,Shao:2017gwu,Kramer:2021jcw,Zhao:2022vig}.\footnote{A STT example with a massless, NMC scalar field evading pulsar-timing constraints is discussed in Ref. \cite{Mendes:2019zpw}.} These systems constrain the amount of scalarization in NSs since the presence of the scalar field can affect drastically their dynamics in a way that its effects should have been already observed, but this is not the case. 
Thus, such astrophysical systems restrict the amplitude of the scalar field in scalarized NSs, as well as the coupling constant whose value controls the deviation of STT from GR. A vanishing NMC constant leads to a theory with a minimally coupled scalar field to gravity, which is of no relevance if the value of the scalar field vanishes cosmologically. 
Binary-system tests and the resulting bounds on STT rely strongly on the range of the scalar field. Since the phenomenon of SC was originally discovered in STT with a massless scalar field $\phi$, which becomes long-ranged after the transition to SC takes place in the star, with an asymptotically falloff as $\sim 1/r$ from its center, $r=0$, its effects can influence the companion in a binary system. Nevertheless, recent analysis by several authors \cite{PhysRevD.92.124016,Ramazanoglu:2016kul,Yazadjiev:2016pcb,Doneva:2016xmf,Sperhake:2017itk,Staykov:2018hhc,Rosca-Mead:2020bzt,Kuan:2023hrh,Lam:2024wpq} show that the SC phenomenon can appear also if one adds a mass term to the STT. The presence of that term results in a short-ranged scalar field (depending on the value of the mass $\mu$) with an exponential decay of Yukawa-type, $\sim e^{-\mu r}/r$, away from the NS, which in turn, may avoid the bounds imposed by observed binary systems.

In a previous work \cite{Degollado:2020lsa}, we performed a numerical analysis of the transition to SC in NSs using polytropic equations of state (EOS), similar to the one reported by Novak several years ago \cite{Novak98a}, except that we have considered a specific STT formulated in the JF with a NMC function quadratic in $\phi$.
Furthermore, we analyzed the effects of the NMC function on the final amplitude of the scalar field after the transition to SC, and confronted its value with current bounds imposed by the binary systems alluded above.

In this paper, we continue the analysis of Ref. \cite{Degollado:2020lsa} by adding a mass term, as in \cite{PhysRevD.92.124016,Ramazanoglu:2016kul,Yazadjiev:2016pcb,Doneva:2016xmf,Sperhake:2017itk,Staykov:2018hhc,Rosca-Mead:2020bzt,Kuan:2023hrh,Lam:2024wpq}, and study the dynamical transition to SC, as well as the range of the scalar field as compared with the typical size of observed binary systems for different values of the field mass $\mu$. We show that some of the observational bounds may be avoided even with large values of the NMC constant, provided that the scalar field mass is sufficiently large. The latter is, however, orders of magnitude smaller than the masses of particles (fundamental or composite) associated with the standard model of particle physics.

This paper is organized as follows. In Sec.~\ref{sec:STT} we specify a STT in the JF with a massive scalar field, and write down the corresponding field equations. We describe static, spherical NSs modeled by a perfect fluid and a polytropic EOS in Sec.~\ref{sec:static_solutions}. In Sec.~\ref{sec:initial_data}, we set initial data for the nonlinear dynamical transition to SC, which is described in Sec.~\ref{sec:dynamics}. 
We discuss our results in Sec.~\ref{sec:discussion}, and conclude in Sec.~\ref{sec:conclusions}. We have crosschecked our numerical results using two different codes, one based on the JF \cite{Degollado:2020lsa}, and one based on the EF \cite{Mendes:2016fby}, thus we summarize relations between the JF and EF in Appendix \ref{app:einstein_frame}. The evolution equations implemented in the latter case are provided in Appendix~\ref{app:evolution_eqns}.
We use units such that the speed of light in vacuum, $c$, reduced Planck's constant, $\hbar$, and Newton's constant, $G$, are all set to one.

\section{Scalar Tensor Theory}
\label{sec:STT}
We shall study a NMC scalar field within the framework of a STT in the JF, as described in \cite{Alcubierre:2010ea, Ruiz:2012jt}. The action of the theory is given by
\begin{eqnarray}
S[g_{ab};\phi;\psi_{\rm m}] &=&
\int \left[\frac{1}{2}f(\phi) R
- \frac{1}{2}(\nabla\phi)^2 - V(\phi)\right] \sqrt{-g} \: d^4x \nonumber \\
&&+ S_{\rm m} [g_{ab}; \psi_{\rm m}]\; ,
\label{eq:action}
\end{eqnarray}
where $S_{\rm m} [g_{ab}; \psi_{\rm m}]$ represents the action for matter fields $\psi_{\rm m}$ (which for the problem at hand are taken as a perfect fluid), and $f(\phi)$ is a positive definite NMC function of the form $f(\phi) = (1 +  \kappa \xi \phi^2)/\kappa \,$, where $\kappa=8\pi$, and $\xi$ is a positive, dimensionless constant that parametrizes deviations from GR. The scalar field potential $V(\phi)$ will be specified below.

Variation of the action~(\ref{eq:action}) with respect to the metric yields field equations that can be recast as effective Einstein field equations,
\begin{equation}
R_{ab}-\frac{1}{2}g_{ab}R = \kappa T_{ab} \; , 
\label{eq:Einst}    
\end{equation}
where the total EMT is
$T_{ab} = \left( T_{ab}^{{\rm fluid}} + T_{ab}^{\phi} + T_{ab}^f \right)/(\kappa f)$,
where contributions from the fluid, scalar field, and function $f$, are respectively given by
\begin{align}  
T_{ab}^{\rm fluid}&= (\epsilon  + p) u_{a}u_{b} + pg_{ab}\; , \label{eq:Tabfluid}\\
T_{ab}^{\phi} &=  (\nabla_a \phi)(\nabla_b \phi) - g_{ab}\left[ \frac{1}{2}(\nabla \phi)^2 + V(\phi) \right ] \; , \label{eq:Tabphi}\\
T_{ab}^f &= \nabla_a \left( f^\prime 
\nabla_b\phi\right) - g_{ab}\nabla_c \left(f^\prime 
\nabla^c \phi\right),\label{eq:TabF}
\end{align}
where $u^a$ is the fluid's 4-velocity, $\epsilon$ is the total energy-density in the rest frame of the fluid, and $p$ is the pressure as measured in the same frame. Here, primes indicate derivatives with respect to the scalar field $\phi$. The diffeomorphism invariance of STT leads to the conservation of the EMT of matter alone, $\nabla_a T^{ab}_{\rm fluid}=0$, which in turn provides the hydrodynamic equations for the fluid.

In this work, we focus on a potential for a massive, but otherwise free, scalar field of the form
\begin{equation}
    V(\phi) = \frac{1}{2}\mu^2\phi^2 \; ,
\end{equation}
where $\mu$ is the mass associated with the scalar field particle. We consider this mass to be in the range $1.33673 \times 10^{-12}$ eV $\leq \mu \leq 1.06859\times$ {$10^{-10}$} eV. For convenience, we operate with a dimensionless mass $\mu/\mu_c$, where $\mu_c=1.33573 \times 10^{-10}$ eV. Then, our interval becomes $0.01\leq \mu/\mu_c \leq 0.8$ (see Table~\ref{t:tablemass}). Moreover, to simplify the notation, we shall rename the dimensionless mass $\mu/\mu_c \to \mu$, except for Sec.~\ref{sec:discussion}.

\begin{table}\centering
  \caption{Equivalence of dimensionless field mass $\mu/\mu_c$, denoted $\mu$ [adim].}
  \begin{tabular}{c| c}
    \hline\hline
    $\mu~[\rm{adim}]$ & $\mu~[\rm{eV}]$\\
    \hline
0.1 & $1.33573\times 10^{-11}$   \\
0.2 & $2.67146\times 10^{-11}$   \\
0.3 & $4.00719\times 10^{-11}$   \\
0.4 & $5.34292\times 10^{-11}$   \\
0.5 & $6.67865\times 10^{-11}$   \\
0.6 & $8.01438\times 10^{-11}$   \\
0.7 & $9.35011\times 10^{-11}$   \\
0.8 & $1.06859\times 10^{-10}$ \\
    \hline \hline
  \end{tabular}\qquad
  \label{t:tablemass}
\end{table}

Regarding the NMC constant $\xi$, although the constraints imposed by pulsars and binary systems yield the bound $\xi \leq 2.5$ for the massless model \cite{Freire2012}, here we consider two values for this parametes, $\xi = \{15, 30\}$, so to enhance the scalarization process and check that the mass term suppresses the range of the scalar field to the extent that the observational constraints are to be avoided. These results are discussed in Sec.~\ref{sec:discussion}.

Variation of the action (\ref{eq:action}) with respect to the scalar field $\phi$ gives the equation of motion
\begin{equation}
g^{ab}\nabla_a\nabla_b \phi + \frac{1}{2}f^\prime R = V^\prime\; .
\label{eq:KGo}
\end{equation}
In the particular case of $f\equiv 1/\kappa$, which in the current scenario corresponds to $\xi \equiv 0$, the theory reduces to Einstein's GR with a minimally coupled scalar field, and with ordinary matter fields. On the other hand, when $\phi\equiv 0$, the STT reduces to GR with a matter content as defined within $S_{\rm m}$ as part of the total action (\ref{eq:action}). The STT can be recast into a Brans-Dicke (BD) type from the action (\ref{eq:action}) by defining a new scalar field $\Phi= f(\phi)$, and by performing the necessary steps so that the kinetic term for the scalar field $\Phi$ takes the desired form with a prefactor that includes a BD function $\omega_{BD} (\Phi)$ \cite{Salgado06}. In the original BD theory, $\omega_{BD}=const$, which has to be of the order of $4\times 10^4$ within the solar system in order to pass the usual tests \cite{Bertotti:2003rm}. 
This bound is important when the scalar field is nonzero within the solar system. However, here we will be concerned only with a non zero scalar field in the neighborhood of a NS that can be part of a binary system, which is otherwise far enough from the solar system to produce noticeable effects on it. In other words, we assume that if the SC phenomenon takes place in a NS, the latter is far enough from the solar system so that, for all practical purposes, the scalar field generated by the star has reached its asymptotic vanishing value in the neighborhood of the solar system, and thus the usual gravity tests remain unaffected as if $\omega_{BD}\rightarrow \infty$.

In the following, we study the consequences of the above theory on spherically symmetric NS models which, for simplicity, we describe in terms of a perfect fluid with a polytropic EOS (see Sec.~\ref{sec:static_solutions}).

\subsection{Scalarization in massive scalar-tensor gravity}
\label{sec:SC}
While SC with a massive scalar field was studied in Refs. \cite{PhysRevD.92.124016,Ramazanoglu:2016kul,Yazadjiev:2016pcb,Doneva:2016xmf,Sperhake:2017itk,Staykov:2018hhc,Rosca-Mead:2020bzt,Kuan:2023hrh,Lam:2024wpq}\footnote{A review can be found in Sec. III-A-2 of Ref. \cite{Doneva:2022ewd}.} with polytropic and other EOS, as far as we are aware, the dynamical transition from unscalarized to scalarized stars was not analyzed. Some of these works study static and spherically symmetric \cite{Ramazanoglu:2016kul}, or rotating NSs \cite{Yazadjiev:2016pcb,Doneva:2016xmf} within a certain class of STT (different from the one we consider here) that supports a nontrivial scalar field that vanishes asymptotically, and under the EF representation.
In principle, those scalarized NSs are supposed to be precisely the result of an evolution like the one we analyze in this paper. Nevertheless, this is only so when the mass of the scalarized NS is below the maximum mass allowed in GR, \textit{i.e.}, in the absence of the scalar field. By definition, a scalarized star corresponds to a preferred lower-energy (\textit{i.e.} lower gravitational mass) configuration than the unscalarized one with fixed total baryon mass. Thus, the final quasistatic configuration after scalarization ensue must have a mass lower than the mass of the initial, unscalarized star (the mass difference being radiated away in the form of scalar field). Therefore, scalarized NSs with masses larger than the maximum mass in GR cannot be the result of an evolution similar to the one considered here, but one of a much more complicated nature that requires perhaps some sort of accretion of scalar field by the star's surroundings, like the accretion of a similar scalar field of cosmic origin, for instance \cite{Degollado:2020lsa}. It would be very interesting in the future to devise a physical scenario that leads to the formation of ``super'' massive scalarized stars, \textit{i.e.} stars with masses larger than the maximum mass models in GR.

\section{Static stellar models}
\label{sec:static_solutions}
Understanding the dynamical transition to scalarization requires first the analysis of static, equilibrium stellar models allowed within the STT under consideration. In Sec.~\ref{sec:initial_data}, some of these stars will be taken as initial data for time evolution. Some others are expected to be the endpoint of evolution after a sufficiently long time.

Spherical star models in hydrostatic equilibrium are solutions of the structure equations that generalize the well known Tolman-Oppenheimer-Volkoff (TOV) equations from GR, with a perfect fluid given by the energy-momentum tensor of Eq.~(\ref{eq:Tabfluid}). Our numerical algorithm solves the structure equations in a formulation that combines the JF fluid variables, and the EF metric $\tilde{g}_{ab}$, which is related to the JF metric $g_{ab}$ by the transformation $g_{ab} = a(\tilde{\phi})^2 \tilde{g}_{ab}$, with a function $a$ that depends on the scalar field $\tilde{\phi}$, defined in the EF.\footnote{A similar Einstein-Jordan hybrid algorithm has been implemented in \cite{Ma:2023sok}.} The relation $\phi(\tilde{\phi})$, the specific function $a(\tilde{\phi})$, and the potential $\tilde{V}(\tilde{\phi})$, are all determined by the conformal transformation between frames, as we outline in Appendix~\ref{app:einstein_frame}. The EF line element in Schwarzschild-like coordinates is given by
\begin{equation}\label{eq:EF_line_element}
d\tilde{s}^2 = - \tilde{N}(r)^2 dt^2 + \tilde{A}(r)^2 dr^2 + r^2 \left(d\vartheta^2 + \sin^2\vartheta d\varphi^2\right).
\end{equation}
The static limit of the evolution system~(\ref{eq:flux-conservative-form})--(\ref{eq:lapse_condition}) results in the following structure equations,\footnote{Under appropriate notation and convention changes, this set of equations reduces to the one in Ref.~\cite{Yazadjiev:2016pcb} in their static limit.}
\begin{align}
 \frac{d\tilde{m}}{dr} &= 4\pi r^2 a^4 \epsilon + \frac{r}{2} (r-2\tilde{m}) \left(\frac{d\tilde{\phi}}{dr}\right)^2 + \frac{r^2}{4}\tilde{V}(\tilde{\phi}), \label{eq:dm}\\
 \frac{d}{dr}\ln \tilde{N} &= \frac{4\pi r^2 a^4 p}{r - 2\tilde{m}} +\frac{r}{2} \left(\frac{d\tilde{\phi}}{dr}\right)^2 + \frac{\tilde{m}}{r(r-2\tilde{m})} \nonumber \\
 &- \frac{r^2}{4(r-2\tilde{m})}\tilde{V}(\tilde{\phi}), \label{eq:dN} \\
 \frac{d^2\tilde{\phi}}{dr^2} &= \frac{4\pi r a^4}{r-2\tilde{m}} \! \left[ \alpha (\epsilon - 3p) + r (\epsilon - p) \frac{d\tilde{\phi}}{dr} \right ]\! \nonumber \\
 &-\left[\frac{2(r-\tilde{m})-r^3\tilde{V}(\tilde{\phi})/2}{r(r-2\tilde{m})} \right]\frac{d\tilde{\phi}}{dr} +\frac{r}{4(r-2\tilde{m})}\frac{d\tilde{V}}{d\tilde{\phi}}, \label{eq:dphi} \\
 \frac{dp}{dr} &= -(\epsilon + p) \left[ \frac{d}{dr}\ln \tilde{N} + \! \alpha \frac{d\tilde{\phi}}{dr} \right], \nonumber \\ \label{eq:dp}
\end{align}
with $\alpha := d \ln{a(\tilde{\phi})}/d\tilde{\phi}$, and the mass aspect function $\tilde{m}$ defined through $\tilde{A}(r)=[1-2\tilde{m}(r)/r]^{-1/2}$.
To close the above system of equations, we specify a polytropic EOS,
\begin{equation}  
 p(\rho) = K \rho_0 \left(\frac{\rho}{\rho_0}\right)^{\gamma} \label{eq:EOS} ,
\end{equation}
where $K$ and $\gamma$ are dimensionless constants, $\rho$ is the fluid's baryon-mass density measured by comoving observers, and $\rho_0$ is a characteristic baryon-mass density given by $\rho_0 = m_b n_0$, where $m_b =1.66 \times 10^{-24}$\;g is the mean baryon mass, and $n_{0}=0.1$ fm$^{-3}$ is the mean baryon-number nuclear density. The total energy-density in this case is given by
\begin{equation}  
 \epsilon(\rho) = \rho + \frac{p}{\gamma-1} \ .
\end{equation}
In this work, we consider a polytrope corresponding to a soft EOS, with $\gamma=2.46$, and $K= 0.0093$. We notice, however, that other polytrope-parameter choices do not change our results qualitatively.

We employ a fourth-order Runge-Kutta algorithm to numerically integrate the structure equations~(\ref{eq:dm})-(\ref{eq:dp}), given a central baryon-mass density $\rho_c :=\rho(0)$ [and thus a central pressure through Eq.~(\ref{eq:EOS})], an initial guess for the central scalar field $\tilde{\phi}_c := \tilde{\phi}(0)$, and regularity conditions $\tilde{m}(0)=0$ and $d\tilde{\phi}/dr(0)=0$. The stellar radius, $R_s$, is defined by the condition  $p(R_s) = 0$ up to numerical tolerance. We iterate on $\tilde{\phi}_c$ until a solution is found such that $\tilde{\phi}(r) \to 0$ as $r$ grows large. Our numerical domain typically consists of $5000$ cells, and extends up to an outermost boundary $r_\text{max} \sim 20R_s$.

Finally, the baryonic mass of a star model is given by
\begin{equation}
M_b = \int_0^{R_s} 4 \pi r^2 \rho~a(\tilde{\phi})^3( 1- 2\tilde{m}/r)^{-1/2} dr,
\end{equation}
which is conserved during time evolution. In the present case of asymptotically flat spacetime, the gravitational (total) mass $M$ of a star model is given by the ADM mass. In our spherically symmetric scenario, $M = \tilde{m}(\infty)$, which includes not only the contribution of the fluid, which has compact support, but also the contribution of the scalar field, which in principle, extends up to spatial infinity. However, since our numerical domain extends only up to $r = r_\text{max}$, our approximate expression for the gravitational mass is $M = \tilde{m}(r_\text{max})$. Beyond $r_\text{max}$, the contribution of the scalar field is negligible, and we have verified that, by extending $r_\text{max}$ to a larger value, the change in $M$ is also negligible. Thus, $M = \tilde{m}(r_\text{max})$ provides a very good approximation to the actual value of the gravitational mass. Moreover, in the region where the scalar field is negligible, $M$ takes the same value in both the EF and the JF, as expected.

\begin{figure}[h]
    \centering
    \begin{subfigure}[]{0.4\textwidth}
         \centering
         \includegraphics[width=\textwidth]{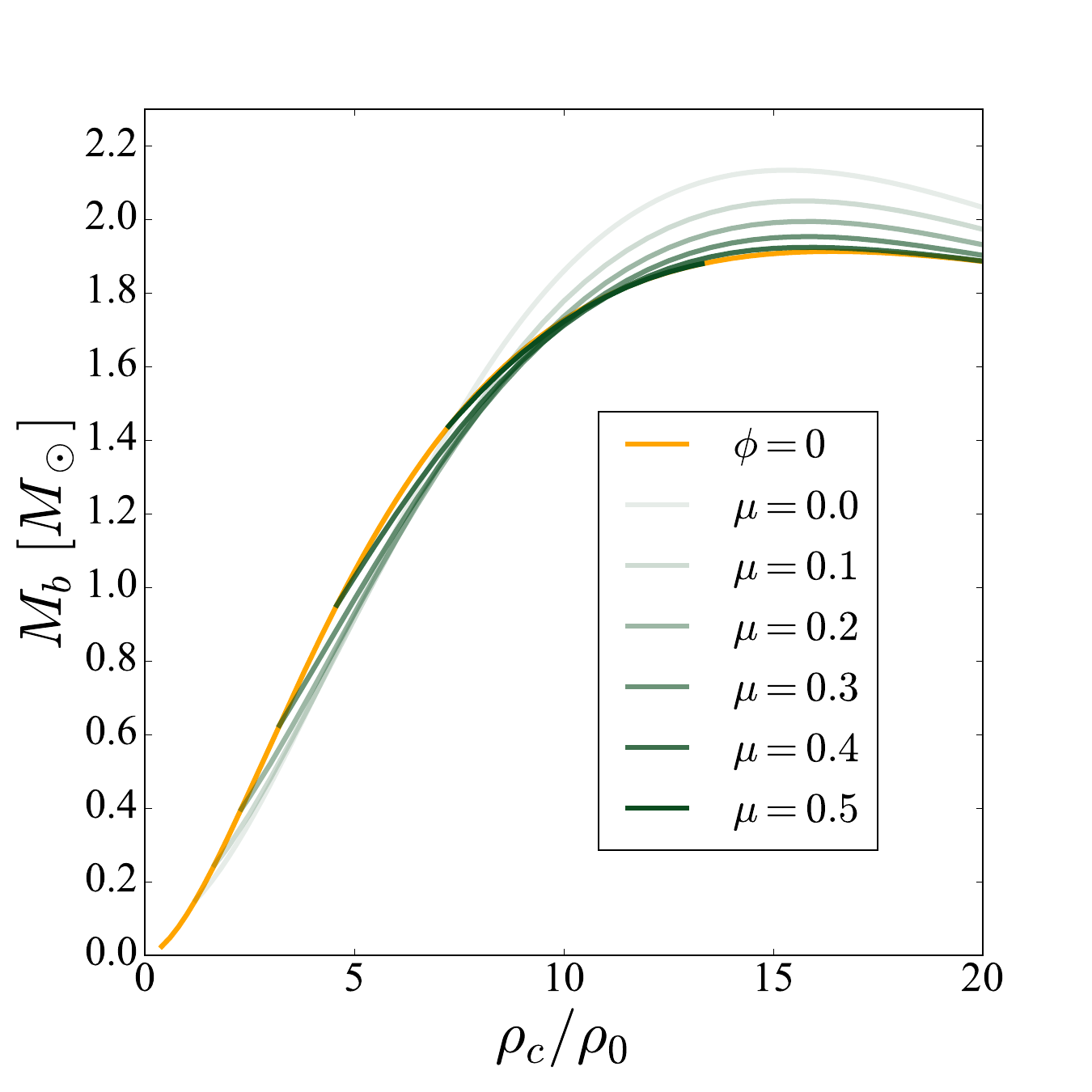}
    \end{subfigure}
    \begin{subfigure}[]{0.4\textwidth}
         \centering
         \includegraphics[width=\textwidth]{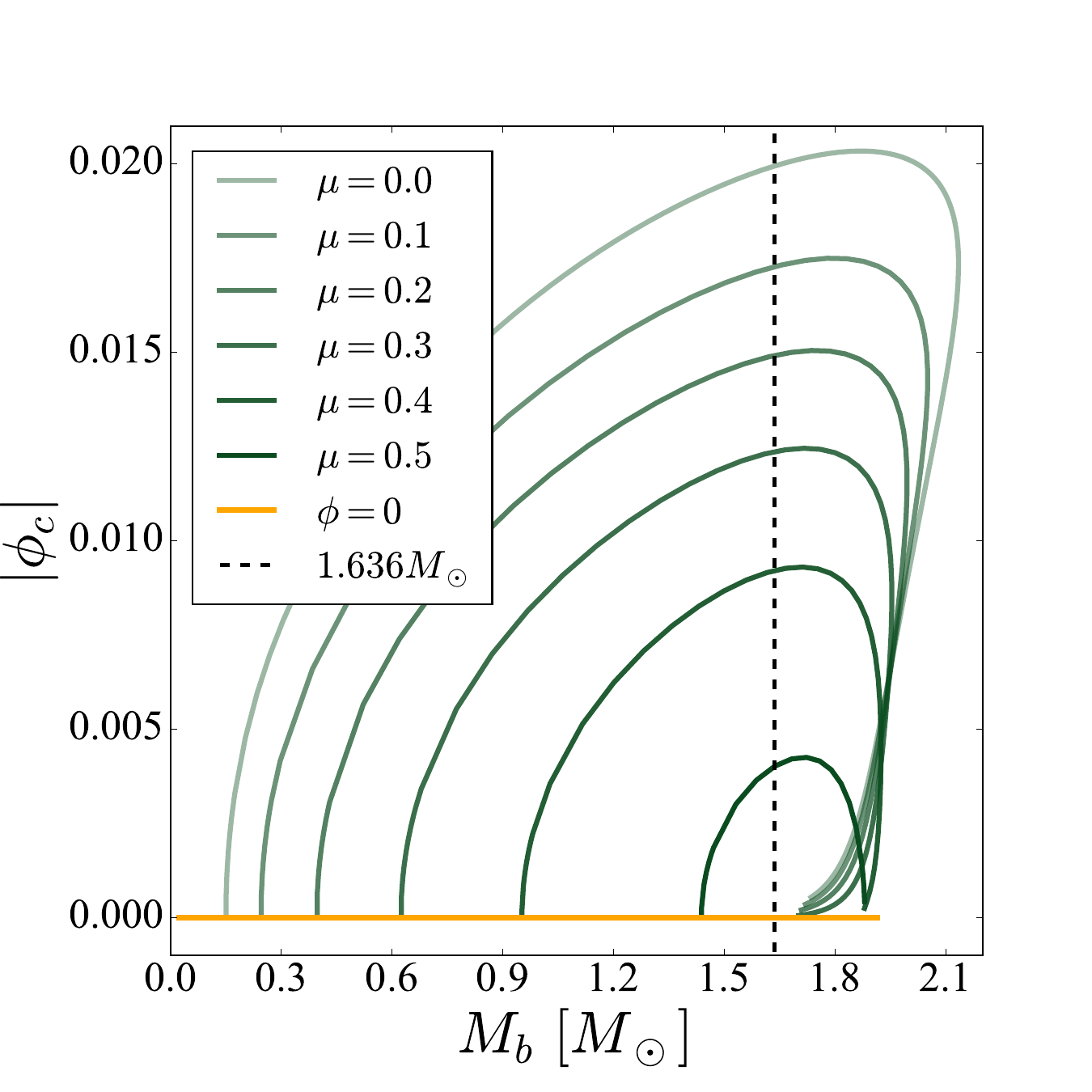}
     \end{subfigure}
        \caption{Sequences of spherical, static stellar models in the case of coupling $\xi=15$. Yellow lines correspond to nonscalarized stars. Green lines correspond to branches of scalarized stars. Top panel: baryon mass {\it vs.} central baryon-mass density. Bottom panel: central value of the scalar field vs. baryon mass. The vertical line corresponds to $M_b = 1.636 M_\odot$, which is the baryon mass of stars undergoing dynamical transition in Sec.~\ref{sec:dynamics}.}
        \label{fig:sequences-xi15}
\end{figure}

Sequences of spherical, static NS models in the case of coupling constant $\xi=15$ are shown in Fig.~\ref{fig:sequences-xi15}, where yellow lines correspond to solutions such that $\phi(r)= 0$, \textit{i.e.} nonscalarized solutions, which are also solutions of the GR equations of structure, thus sometimes referred to as ``GR-like'' solutions. In turn, green lines in Fig.~\ref{fig:sequences-xi15} correspond to spontaneously scalarized solutions, meaning that they possess a nontrivial $\phi(r)$ profile that can extend beyond the stellar surface.
The bottom panel shows the central value of the scalar field, $\phi_c=\phi(0)$, as a function of baryonic mass $M_b$. For this value $\xi=15$, scalarized solutions feature no nodes in their $\phi(r)$ profile, in the sense that its sign never changes. Notice that scalarized solutions reach a maximum baryonic mass larger than the GR-like solutions. Beyond this maximum, stars are unstable to gravitational collapse. Most importantly, in the range where scalarized solutions exist, GR-like solutions are unstable to scalar field perturbations, and the endpoint of this instability is expected to be a scalarized solution of the same baryonic mass. Also notice (specially from the bottom panel of Fig.~\ref{fig:sequences-xi15}) that branches of scalarized solutions---progressively darker green lines---become shorter as the scalar field mass $\mu$ increases. In fact, scalarized solutions exist only up to values of $\mu \lesssim 0.5$, meaning that, for $\mu \gtrsim 0.5$, the only possible scalar-field profile that vanishes asymptotically is the trivial one, $\phi(r)\approx 0$ (within the numerical accuracy considered in this work, and for the range of stellar baryon masses we explored for this particular EOS), like in GR. The nonappearance of spontaneous scalarization for such large $\mu$ is presumably due to the suppression of the scalar field by the Yukawa term, so that the field becomes short ranged to the extent that it disappears completely as one can appreciate in Fig.~\ref{fig:phi_profiles}. The scalar field profiles in that figure (top panel) have central values $\phi_c$ associated with those of Fig.~\ref{fig:sequences-xi15} (bottom panel) that intersect the vertical dashed line. In particular, we see in Fig.~\ref{fig:sequences-xi15} (bottom panel) that, for $\mu \gtrsim 0.5$, such intersection corresponds to $\phi_c=0$, while in Fig.~\ref{fig:phi_profiles} (top panel) we see that the range of the scalar field becomes shorter as $\mu$ increases, and it vanishes completely for $\mu \gtrsim 0.5$ inside and outside the star.
\begin{figure}[h]
     \centering
     \begin{subfigure}[]{0.4\textwidth}
         \centering
         \includegraphics[width=\textwidth]{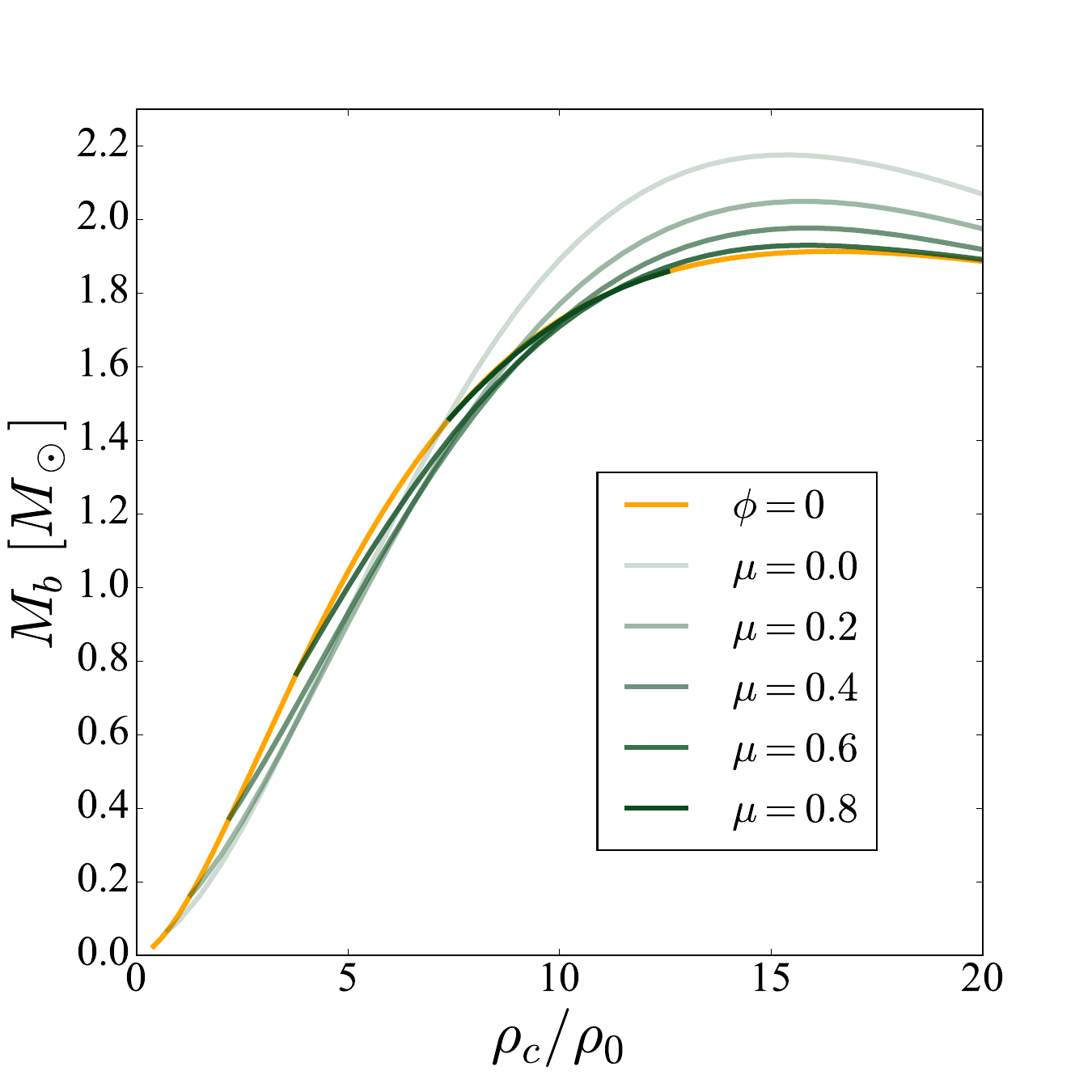}
     \end{subfigure}
     \begin{subfigure}[]{0.4\textwidth}
         \centering
         \includegraphics[width=\textwidth]{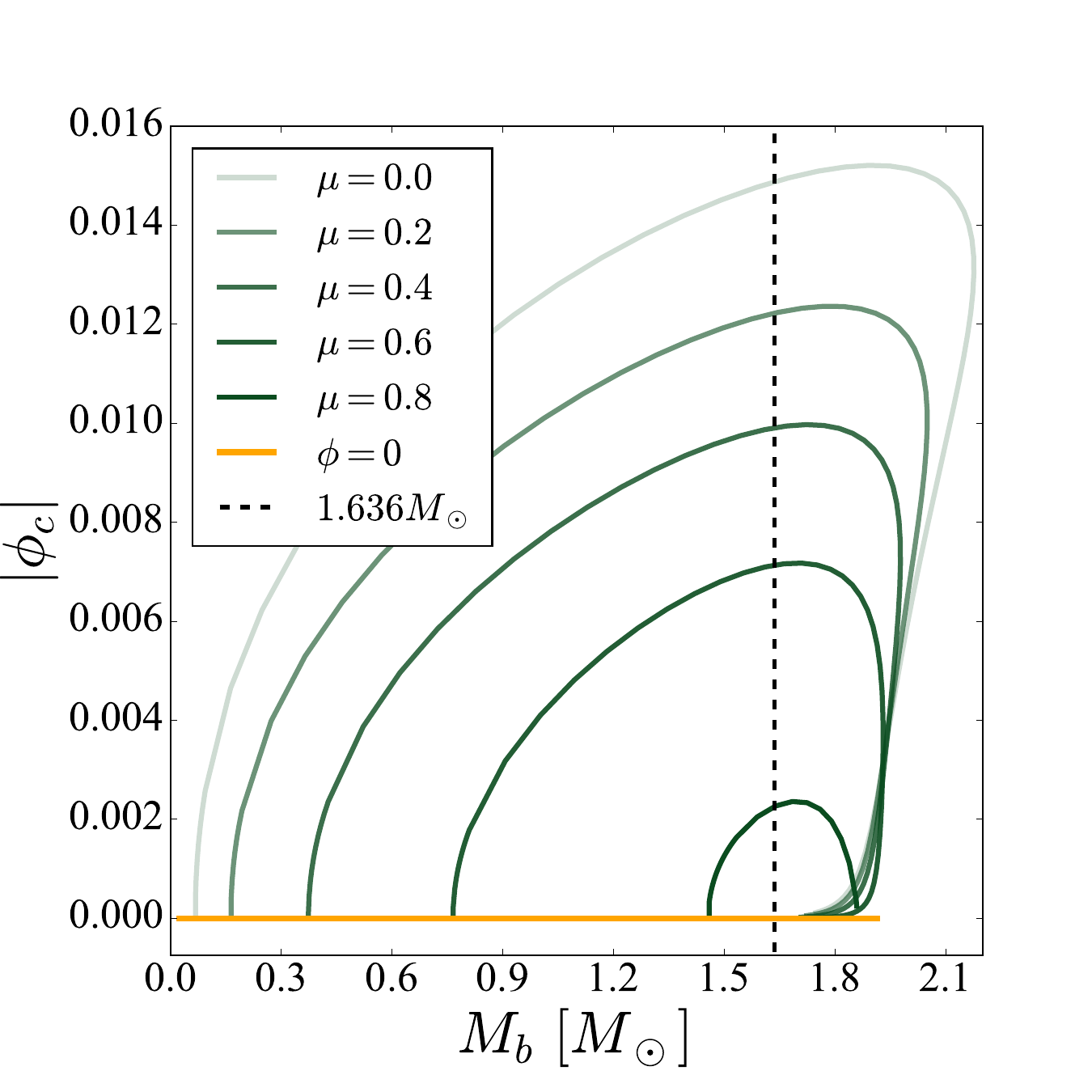}
     \end{subfigure}
        \caption{Same as Fig.~\ref{fig:sequences-xi15}, for the case of coupling $\xi=30$.}
        \label{fig:sequences-xi30}
\end{figure}

\begin{figure}[h]
    \centering
    \includegraphics[width=0.4\textwidth]{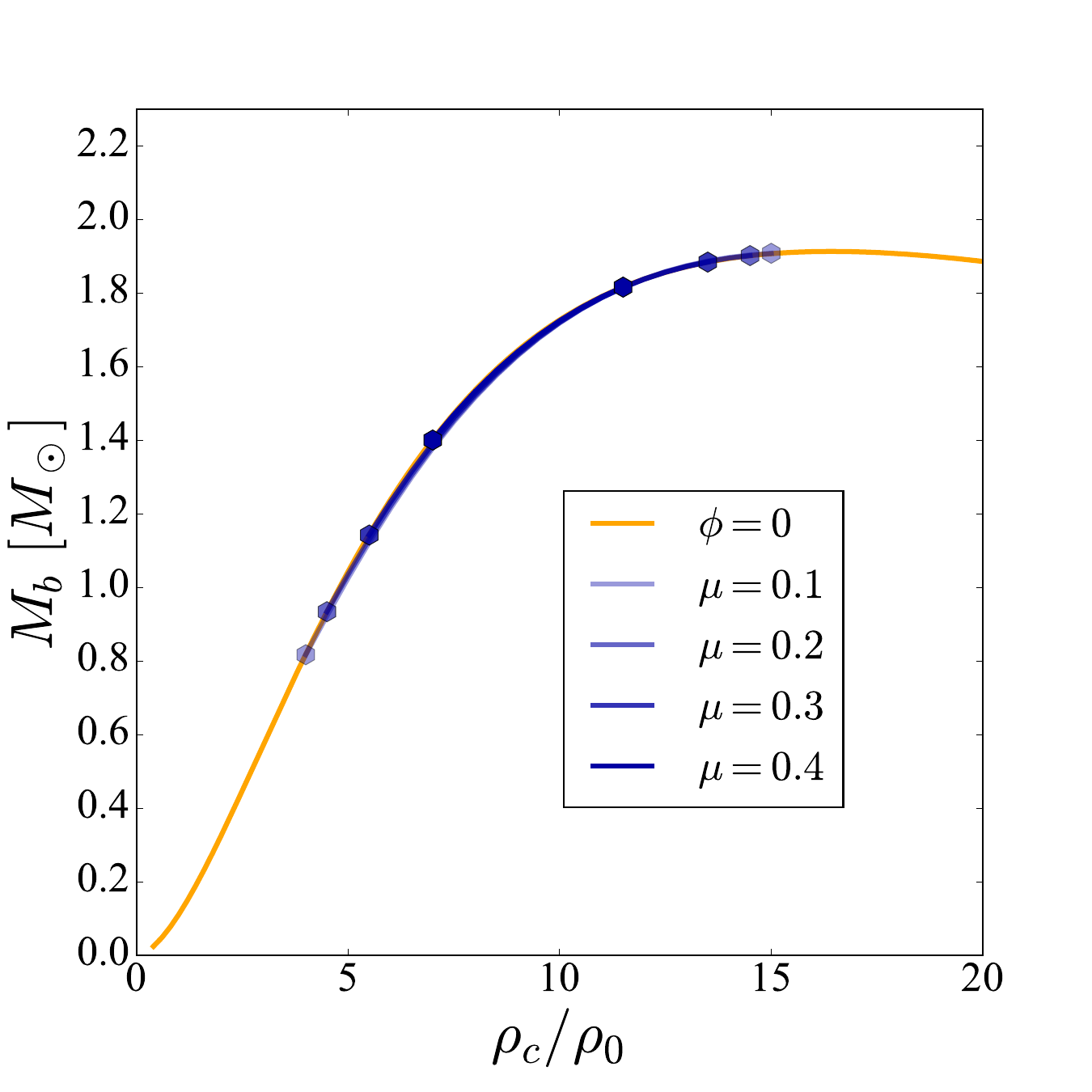}
    \caption{Sequences of static, spherical stellar models in the case of $\xi=30$. The yellow line corresponds to nonscalarized stars. Blue branches correspond to scalarized stars with 1-node in their $\phi(r)$ profile. Dot marks indicate junctions to the nonscalarized (yellow) sequence.}
    \label{fig:sequences-xi30-1n}
\end{figure}

The case of coupling constant $\xi=30$ features the same qualitative characteristics, as can be seen in Fig.~\ref{fig:sequences-xi30}. There are two main differences though. First, in this case, spontaneously scalarized solutions do exist for $\mu > 0.5$, up to $\mu=0.8$. Second, besides the 0-node scalarized solutions---green branches---for $0 < \mu \lesssim 0.4$ there are scalarized solutions whose $\phi(r)$ profile features one node, meaning that it changes sign exactly once. The corresponding branches are shown in Fig.~\ref{fig:sequences-xi30-1n}, where progressively blue lines correspond to progressively higher values of mass $\mu$. It is clear that the larger $\mu$, the smaller the 1-node scalarized branch.

Several questions arise with respect to the existence of scalarized solutions with nodes in their $\phi(r)$ profile. Our numerical experience suggests that there is a critical value in the interval $15< \xi < 30$ such that 1-node solutions show up, and then, the higher $\xi$, the more nodes appear in scalarized solutions. We have considered up to $\xi=240$, and we have found up to 5 nodes in scalar field profiles.
A capital question regarding scalarized solutions with nodes is about their stability to scalar field perturbations. In all of the cases we explored numerically, for a fixed baryonic mass $M_b$, the configuration with no nodes always possesses a gravitational mass $M$ smaller than any configuration with nodes. This suggests that 0-node stars are energetically favored over stars with nodes with the same baryonic mass. We confirm this expectation in Sec.~\ref{sec:dynamics}.

\begin{figure}[h]
    \centering
    \includegraphics[width=0.4\textwidth]{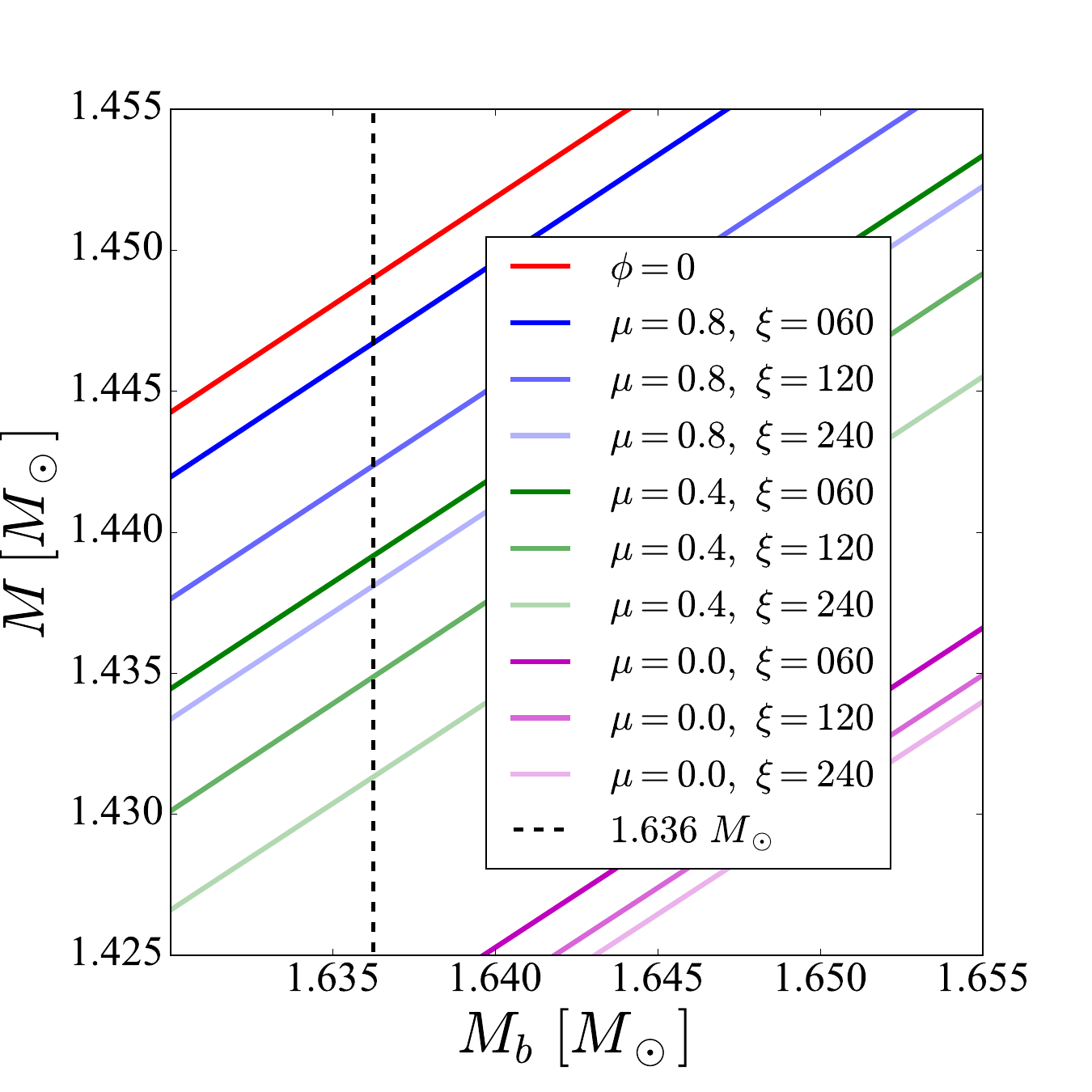}
    \caption{Sequences of 0-node solutions. For any given baryonic mass $M_b$ and fixed scalar field mass $\mu$, the larger the coupling constant $\xi$, the lower the gravitational mass $M$. Likewise, for a given coupling $\xi$, the smaller the field mass $\mu$, the lower the gravitational mass $M$. Therefore, smaller masses $\mu$ and larger couplings $\xi$ lead to energetically preferred scalarized configurations. The vertical line indicates $M_b = 1.636 M_\odot$, which is the baryon mass of stars undergoing dynamical transition in Sec.~\ref{sec:dynamics}}
    \label{fig:M_vs_Mb-0n}
\end{figure}

A more complete picture of what makes a 0-node scalarized star energetically preferred with respect to a nonscalarized one, in terms of its gravitational mass $M$, is summarized in Fig.~\ref{fig:M_vs_Mb-0n}. It reveals that, for a given baryonic mass $M_b$ and coupling $\xi$, a smaller field mass $\mu$ corresponds to smaller gravitational mass $M$. Most remarkably, for fixed $M_b$, and given $\mu$ and $\xi$, the nonscalarized solution ($\phi(r)=0$, red line) is the less energetically favored.

\begin{figure}[h!]
     \centering
     \begin{subfigure}[]{0.4\textwidth}
         \centering
         \includegraphics[width=\textwidth]{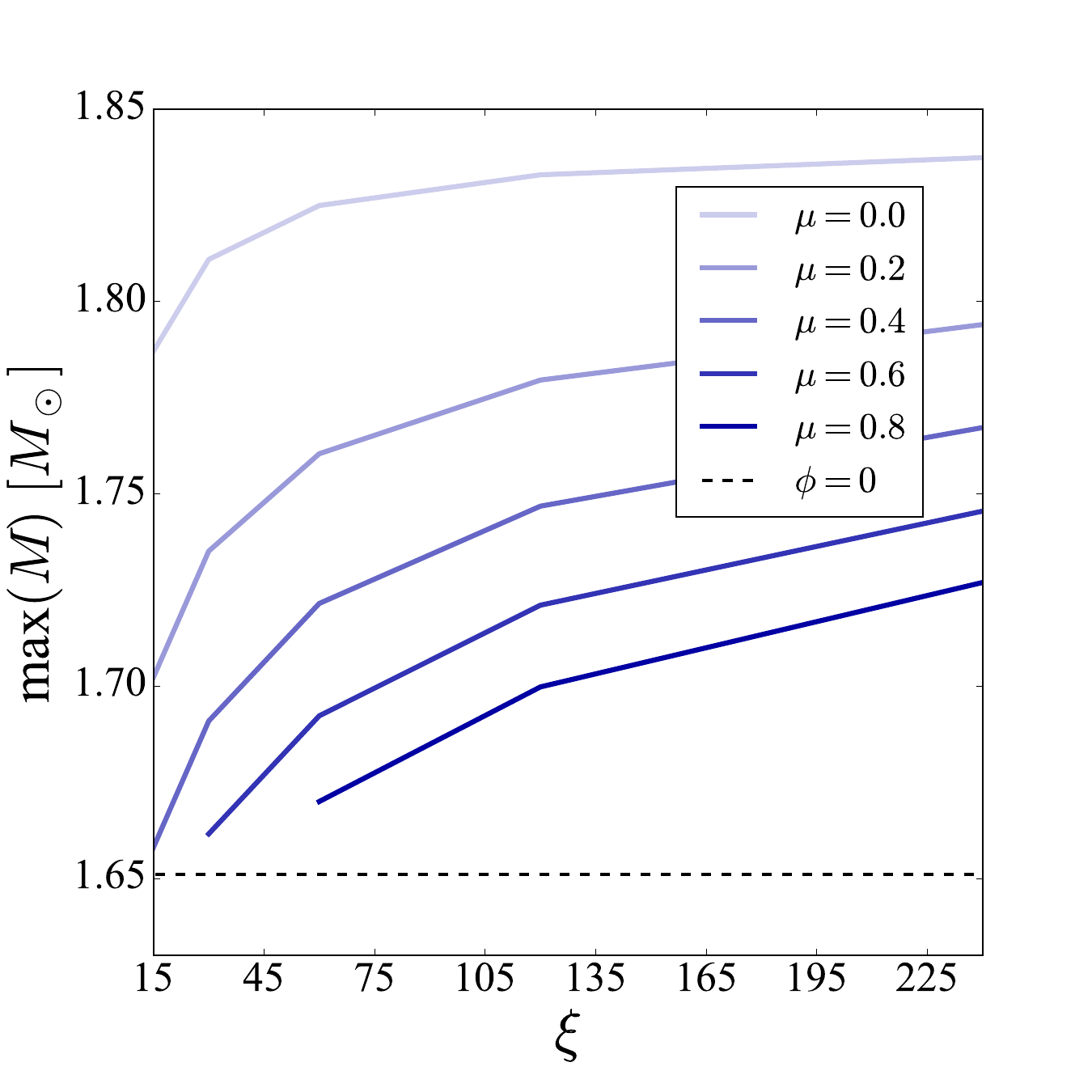}
     \end{subfigure}
     \begin{subfigure}[]{0.4\textwidth}
         \centering
         \includegraphics[width=\textwidth]{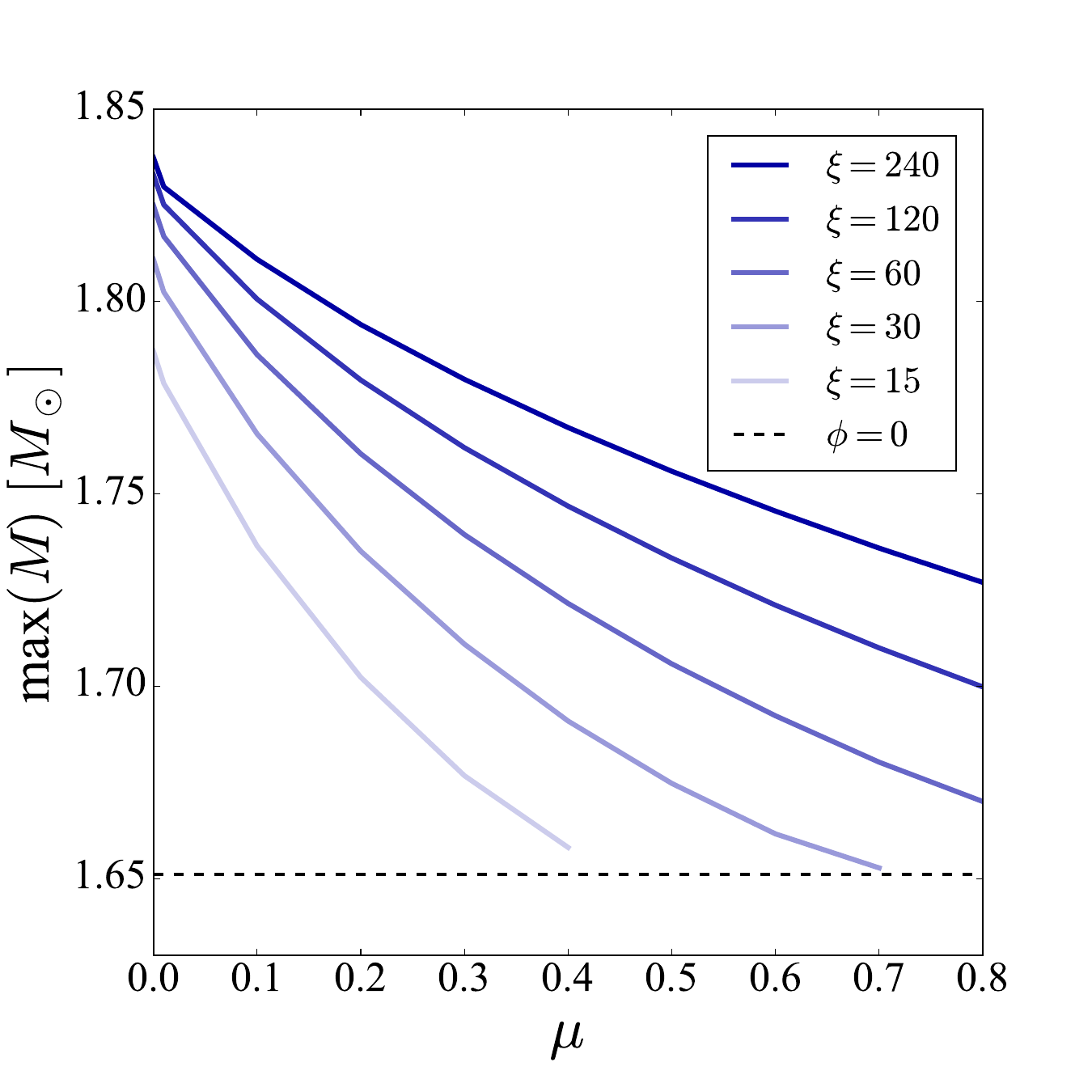}
     \end{subfigure}
        \caption{Maximum of the gravitational mass $M$ as a function of coupling constant $\xi$ (top panel) and scalar field mass $\mu$ (bottom panel). Horizontal dashed lines indicate the maximum mass allowed for NSs in GR, for the EOS under consideration.}
        \label{fig:max_M}
\end{figure}

Figure~\ref{fig:max_M} provides complementary aspects of static, scalarized solutions with no nodes. It shows that, for fixed coupling $\xi$, smaller field masses $\mu$ allow for scalarized stars to have a larger maximum gravitational mass $M$.
The lack of points in the case of $\xi=15$ for $\mu=\{0.6, 0.8\}$ is due to nonexistence of scalarized solutions beyond $\mu=0.5$ for such a small $\xi$. Notably, keeping the EOS fixed, the maximum mass of scalarized solutions is always larger than the maximum mass allowed for nonscalarized solutions, represented by a dashed line in Fig.~\ref{fig:max_M}.

\section{Initial data}
\label{sec:initial_data}

\begin{table*}[]
  \centering
 \begin{tabular}{c |c |c | c | c | c | c } 
 \hline\hline
 Index & Branch & Color in Fig.~\ref{fig:initial_data_for_fixed_1.4Mb} & $\rho_c/\rho_0$ & $M$ [$M_\odot$] & $R$ [km] & $|\phi_c|$ \\
 \hline
1 & $\phi = 0$ & Black & $6.9851$ & $1.26385$ & $10.56$ & $0$ \\
 \hline 
2 & Scalarized, $0$ nodes & Blue & $7.5237$ & $1.25722$ & $10.53$ & $0.0105$ \\
\hline
3 & Scalarized, $1$ node & Green & $7.0579$ & $1.26384$ & $10.56$ & $0.0029$ \\ 
 \hline\hline
\end{tabular}
  \caption{Parameters of static, stellar models with fixed baryonic mass $M_b = 1.4 M_\odot$, in the case of $\xi=30$, $\mu=0.3$. The scalarized solution with no nodes in its scalar-field profile minimizes the gravitational mass $M$, and is therefore energetically favored. Initial data consisting of solutions 1 or 3 are expected to evolve to a configuration consistent with solution 2.}
  \label{tab:comparing_M}
\end{table*}

Our main goal is to investigate the nonlinear dynamical evolution from unscalarized NSs to stable, spherical, scalarized stars. This task requires numerical integration of the STT field equations, coupled to the scalar field equation of motion, as well as the hydrodynamical equations. Initial data for time evolution consist of nonscalarized stars in hydrostatic equilibrium, \textit{i.e.} $\phi=0$ solutions of the structure equations~(\ref{eq:dm})-(\ref{eq:dp}), except for a small scalar field perturbation to trigger the transition of the star to the scalarized state. The initial configuration has a gravitational mass below the maximum associated to the given EOS, and is stable against gravitational collapse into a black hole.

\begin{figure}[h]
    \centering
    \includegraphics[width=0.4\textwidth]{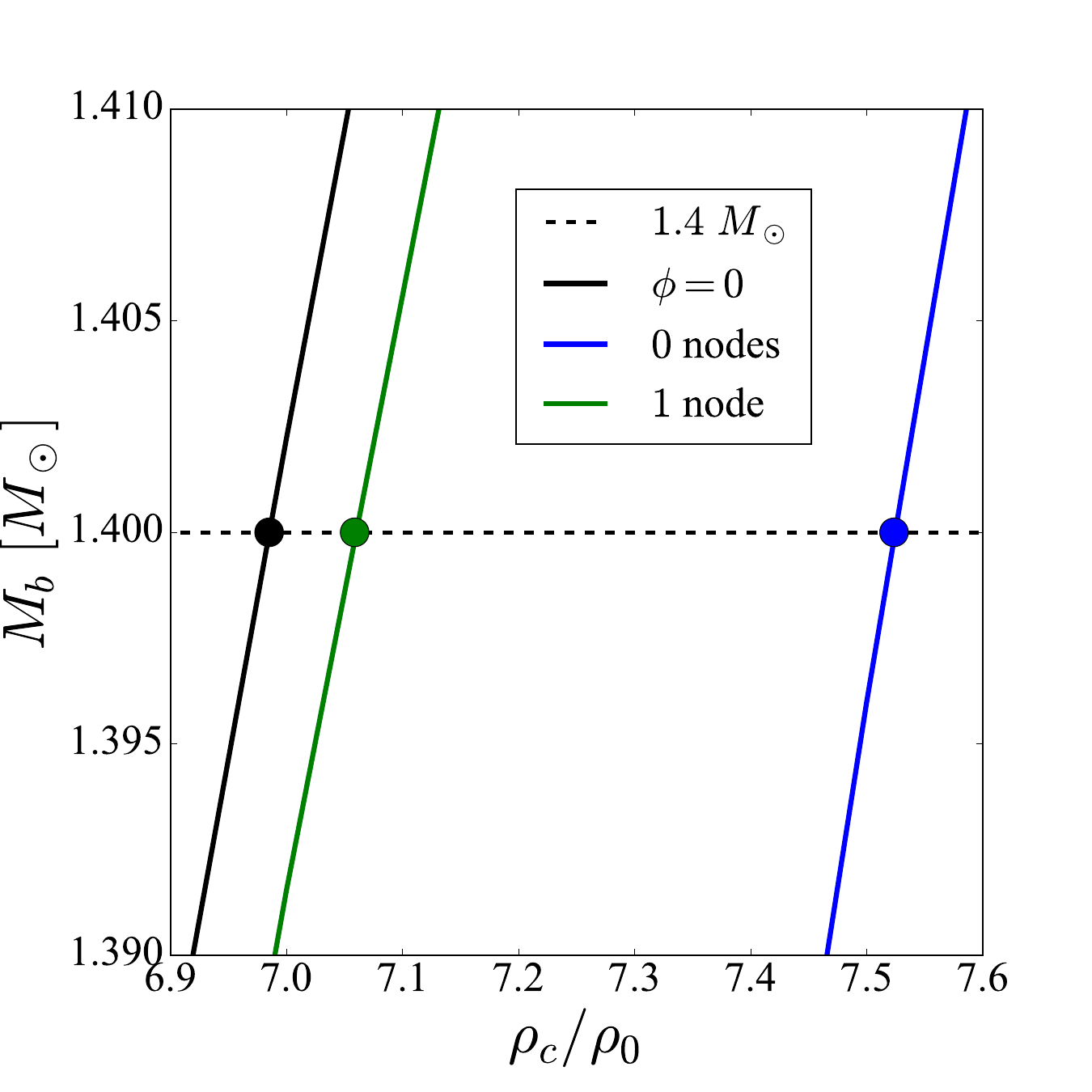}
    \caption{Solid lines are small portions of NS sequences. Highlighted points correspond to $M_b = 1.4 M_\odot$ stars in the case of $\xi=30$, $\mu=0.3$. The 0-node (blue point) configuration is energetically favored among the three highlighted stars, as can be read from Table~\ref{tab:comparing_M}. Thus, if either the black or green points are taken as initial data, then time evolution drives them to a configuration consistent with the blue point configuration. This expectation is confirmed in Fig.~\ref{fig:1_4M-time_evolution}. Our results show that 0-node configurations are always energetically favored over those with one or more nodes.}
    \label{fig:initial_data_for_fixed_1.4Mb}
\end{figure}

\begin{figure}
    \centering
    \includegraphics[width=0.5\textwidth, angle=0]{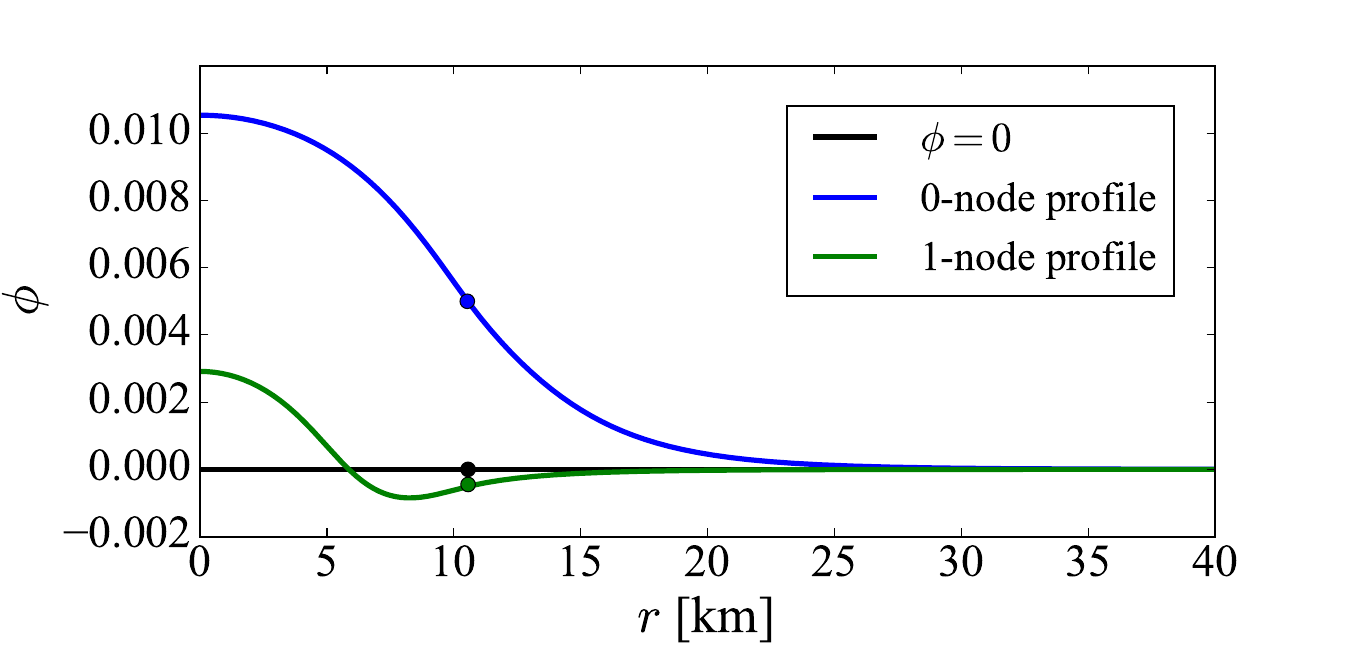}
\caption{Scalar-field radial profiles corresponding to the static solutions described in Table~\ref{tab:comparing_M}, which are associated with colored circles in Fig.~\ref{fig:initial_data_for_fixed_1.4Mb}. The stellar radius in each case is indicated by small circles.}
\label{fig:phi_profiles-14}
\end{figure}

In view of the existence of scalarized solutions with an increasing number of nodes in their $\phi(r)$ profile as the coupling $\xi$ gets larger, a natural question is whether or not slightly perturbed initial data are expected to evolve to a scalarized configuration with nodes. To give the answer, we focus on a scenario where scalarized 1-node solutions exist, concretely the case of $\xi=30$, $M_b = 1.4M_\odot$, and $\mu=0.3$. Figure~\ref{fig:initial_data_for_fixed_1.4Mb} shows small portions of branches of static, equilibrium solutions in this case. The black line is a sequence of nonscalarized stars, whereas blue and green lines correspond to 0- and 1-node sequences of scalarized stars, respectively. We consider the three existing solutions with baryon mass $M_b = 1.4 M_\odot$. We label them with dots in Fig.~\ref{fig:initial_data_for_fixed_1.4Mb}, show their $\phi$-profiles in Fig.~\ref{fig:phi_profiles-14}, and report their parameters in Table~\ref{tab:comparing_M}, where it is clear that the 0-node scalarized solution (blue dot) features the smallest gravitational mass $M$. Therefore, among the three solutions, the 0-node configuration is energetically favored, and it is thus expected to be the end state of time evolution if one takes initial data consisting on either the nonscalarized solution (black dot), or the 1-node solution (green dot). In other words, since the dynamics ideally keeps the baryon mass constant, then starting either in the black or green dots of Fig.~\ref{fig:initial_data_for_fixed_1.4Mb}, a small perturbation will onset the scalar field instability whose end-state after sufficiently long time should be consistent with the blue dot configuration. This expectation is confirmed in Sec.~\ref{sec:dynamics}, in particular Fig.~\ref{fig:1_4M-time_evolution}.
According to our numerical inspection, static, scalarized solutions with 0-nodes are always energetically favored over solutions with nodes. This observation motivates the conjecture that the dynamical transition from nonscalarized stars should always lead to scalarized solutions with 0-nodes. This conjecture holds in all of our numerical simulations of dynamical transition to scalarization, as reported in Sec.~\ref{sec:dynamics}.
\begin{figure}[h]
    \centering
    \begin{subfigure}[h]{0.4\textwidth}
         \centering
         \includegraphics[width=\textwidth]{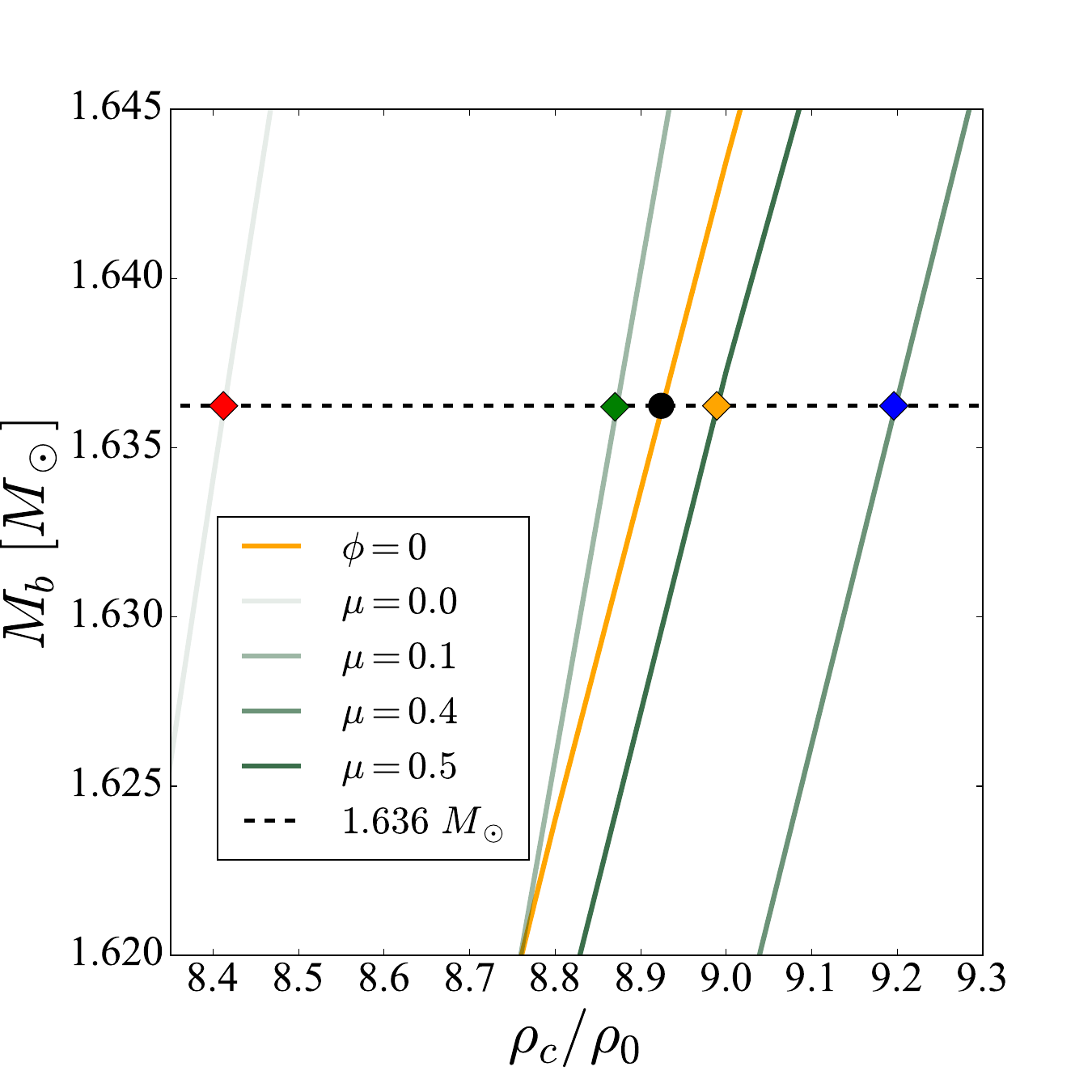}
    \end{subfigure}
    \begin{subfigure}[h]{0.4\textwidth}
         \centering
         \includegraphics[width=\textwidth]{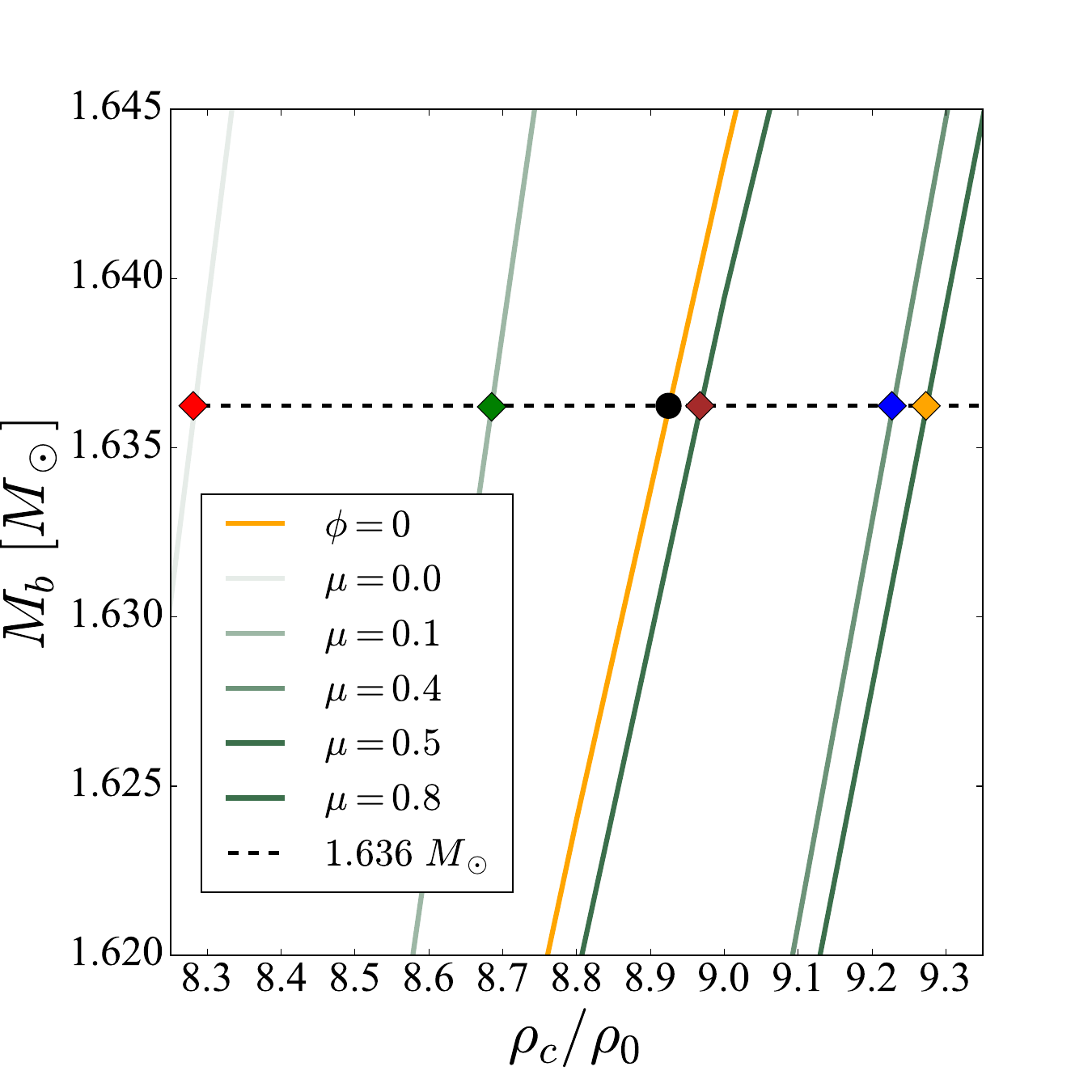}
     \end{subfigure}
    \caption{Top panel: Zoom-in to star sequences of Fig.~\ref{fig:sequences-xi15} ($\xi=15$). Bottom panel: Zoom-in to sequences of Fig.~\ref{fig:sequences-xi30} ($\xi=30$). Black dots represent initial data for the time evolutions shown in Figs.~\ref{fig:time_evolution_xi15}~and~\ref{fig:time_evolution_xi30}, respectively. Colored diamonds indicate expected endpoints of those time evolutions.}
    \label{fig:initial_data}
\end{figure}
Static solutions chosen as initial data for time evolution with different choices of $\mu$ have a central energy-density such that they lie in the branch {\it unstable} to scalar field perturbations. In other words, initial data are stable toward BH formation, but unstable toward transition to a scalarized state. We have arbitrarily chosen initial data with baryonic mass $M_b = 1.636 M_\odot$, which corresponds to $\rho = 8.9245\rho_0$. These initial data are located in the yellow sequences of Figs.~\ref{fig:sequences-xi15}~and~\ref{fig:sequences-xi30}. Indeed, since the central value of the scalar field for initial data is $\phi_c=0$, this corresponds to the intersection of the vertical dashed lines with horizontal, yellow lines of bottom panels of Figs.~\ref{fig:sequences-xi15}~and~\ref{fig:sequences-xi30}. Time evolution drives the initial data to a scalarized configuration with $|\phi_c| \neq 0$, ideally conserving the baryon mass, thus corresponding in those two figures to the intersection of the vertical dashed line with a green curve, depending on the field mass $\mu$.
Another viewpoint of these initial data, in terms of the central value of their baryon-mass density $\rho_c$, is displayed in Fig.~\ref{fig:initial_data}, which is a zoom-in to the top panels of Figs.~\ref{fig:sequences-xi15}~and~\ref{fig:sequences-xi30}. In both cases, $\xi=15$ and $\xi=30$, a black dot indicates initial data for time evolution. Horizontal, dashed lines correspond to constant baryon mass $M_b = 1.636 M_\odot$. Depending on the value of $\mu$, colored marks indicate the expected final state of time evolution, which consist of 0-node scalarized stars. In the next section, concretely in Figs.~\ref{fig:time_evolution_xi15}~and~\ref{fig:time_evolution_xi30}, we show evidence that the dynamics brings initial data to the expected end-states.

\section{Dynamical transition to scalarized states}
\label{sec:dynamics}
The dynamical process of scalarization is given by the solution in space and time of the effective Einstein-scalar-field system (\ref{eq:Einst})--(\ref{eq:KGo}), together with the hydrodynamic equations
\begin{eqnarray}
    \nabla_a T^{ab}_{\rm fluid} &=& 0, \label{eq:Tab_divergence}\\
    \nabla_a (\rho u^a) &=& 0.\label{eq:continuity}
\end{eqnarray}
We solve the whole system in spherical symmetry using two independent numerical codes.

The first code is also the one used to construct the static solutions of Sec.~\ref{sec:static_solutions}. It evolves JF fluid variables, together with the EF metric functions and scalar field. The full evolution system can be found in Appendix~\ref{app:evolution_eqns}. The JF scalar field and metric components are recovered through the relations given in Appendix~\ref{app:einstein_frame}. This code uses a finite-volume method with evolution equations written in conservative form. Details on the formalism, methods, boundary conditions, vacuum treatment, and convergence tests, can be found in Refs.~\cite{Mendes:2016fby} and \cite{Mendes:2021fon}.

\begin{figure}[h]
   \centering
    \includegraphics[width=0.5\textwidth, angle=0]{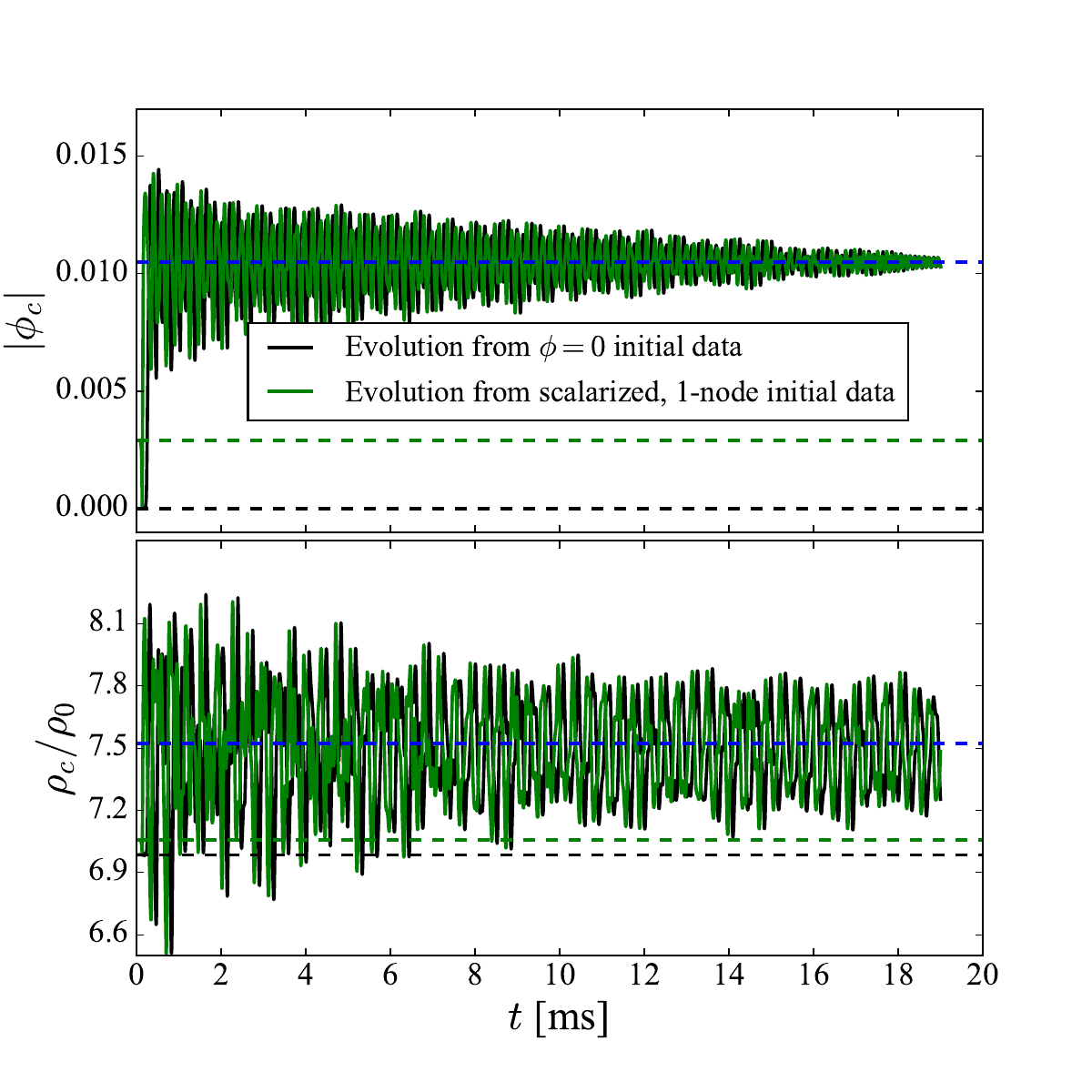}
    \caption{Time evolution of the central value of the scalar field (top panel) and central value of baryon-mass density (bottom panel), in the case of $\xi=30$, $\mu=0.3$. The initial data corresponds to the black and green dots of Fig.~\ref{fig:initial_data_for_fixed_1.4Mb} In both cases, time evolution drives the stars to the 0-node scalarized solution, consistent with the blue dot configuration in Fig.~\ref{fig:initial_data_for_fixed_1.4Mb}, which is energetically preferred, as described in Table~\ref{tab:comparing_M}. Horizontal lines indicate values associated with the initial data and the expected endpoint of evolution.}
    \label{fig:1_4M-time_evolution}
\end{figure}

The second code solves the evolution system entirely in the JF. It uses a modified BSSN formalism adapted to spherical coordinates like in~\cite{Degollado:2020lsa}, with the spatial metric given by
\begin{equation}
ds^2_3 = \psi(t,r)^{4}\Big[a(t,r){dr}^2 
+ r^2 b(t,r)\left(d\vartheta^2 + \sin^2 \vartheta d\varphi^2 \right) \Big] \, .
\end{equation}
This code was used in the past for an almost identical scenario, except that the scalar field was massless \cite{Degollado:2020lsa}, and also when analyzing the dynamical transition to SC in spherically symmetric boson stars \cite{Alcubierre:2010ea,Ruiz:2012jt}. A detailed description of the numerical method and variables employed during time evolution is given in Refs. \cite{Alcubierre:2019qnh, Degollado:2020lsa}.

We have checked that, despite their different approaches and methods, the numerical codes that we used for analysis in the Einstein and Jordan frames, consistently return the same phenomenology, which further supports the equivalence between these two formulations of STT.

Small numerical errors in the initial data construction are enough to onset the instability to scalar field perturbations. Time evolution then goes through an early phase of exponential growth, which should be comparable against predictions from linear perturbation theory.\footnote{To our knowledge, linear perturbation equations are available only in the massless case. See, for example, Ref.~\cite{Mendes:2018qwo}.}
An alternative approach to the onset of the instability is to induce a quicker dynamical transition to scalarization by means of a Gaussian perturbation into the scalar field, assuming staticity on the fluid and spacetime. This approach saves computing time, but misses details of the scalar-field exponential growth at early times, in the small-perturbation regime.

As a first dynamical experiment, we take initial data represented by the black and green dots of Fig.~\ref{fig:initial_data_for_fixed_1.4Mb}, which are associated with a static unscalarized star, and with 1-node static scalarized star, respectively. As anticipated in Sec.~\ref{sec:initial_data}, once the scalar-field instabilities have triggered in both cases, and after sufficiently long times, the dynamical solution settles around the static scalarized 0-node solution represented by the blue dot in Fig.~\ref{fig:initial_data_for_fixed_1.4Mb}, because this is the energetically preferred configuration with the same baryon mass $M_b = 1.4 M_\odot$. The expected behavior is confirmed in Fig.~\ref{fig:1_4M-time_evolution}, where solid lines represent the time evolution of the central value of the scalar field, $\phi_c$ (top panel) and the central value of the baryon-mass density, $\rho_c$ (bottom panel). In both panels, dashed horizontal lines indicate values associated with the static solutions represented by dots in Fig.~\ref{fig:initial_data_for_fixed_1.4Mb}, with the same color code. In both cases, the dynamical solution features a {\it quasistatic} configuration with a gravitational mass lower than the initial one (\textit{i.e.} the initial gravitational mass minus some energy radiated away in the form of scalar field).

\begin{figure}[]
    \centering
    \includegraphics[width=0.5\textwidth, angle=0]{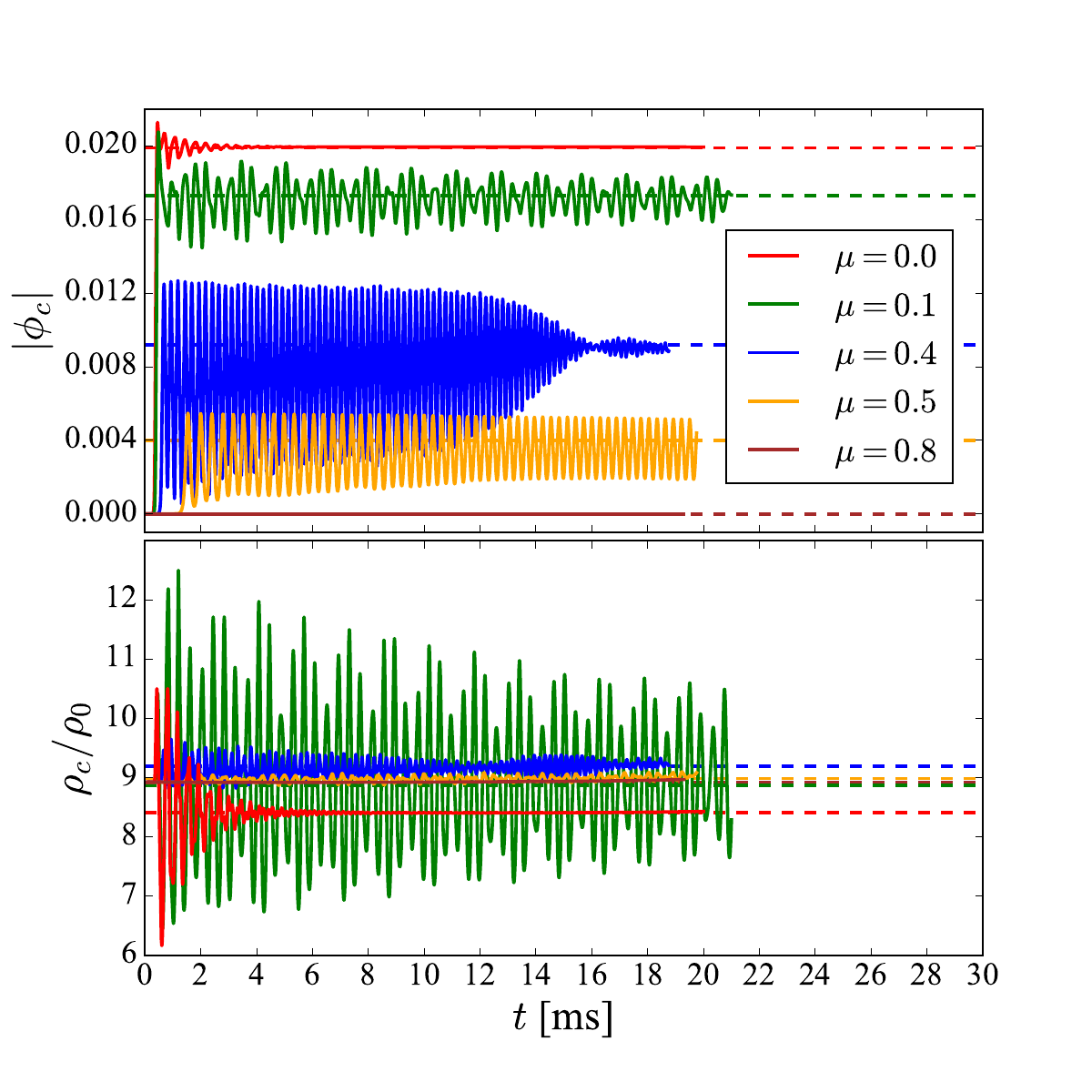}
    \caption{Time evolution of the central value of the scalar field (top), and the central baryon-mass density (bottom), in the case of coupling $\xi = 15$.}
    \label{fig:time_evolution_xi15}
\end{figure}

\begin{figure}[]
    \centering
    \includegraphics[width=0.5\textwidth, angle=0]{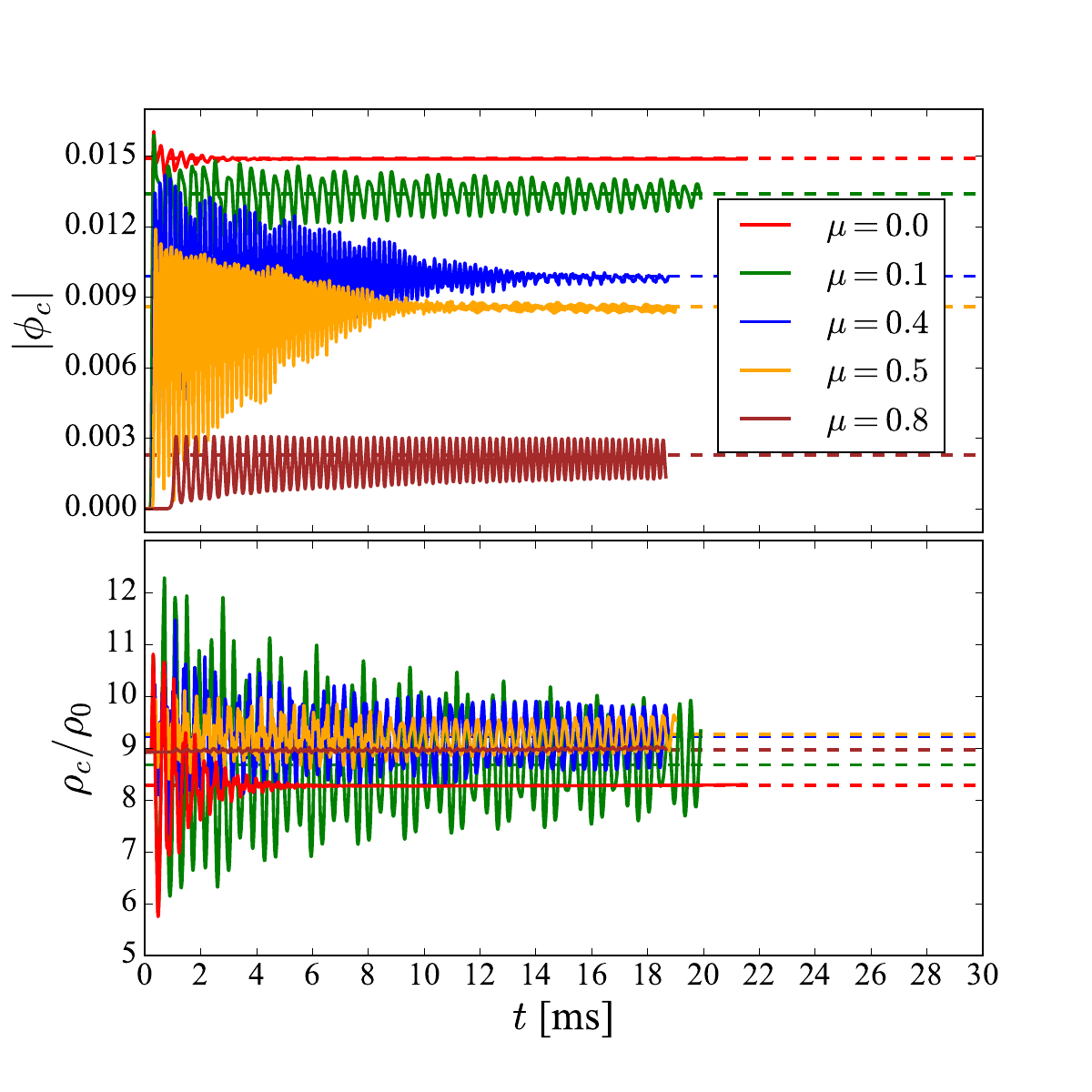}
    \caption{Same as Fig. \ref{fig:time_evolution_xi15} with coupling constant $\xi = 30$.}
    \label{fig:time_evolution_xi30}
\end{figure}

We now discuss the time evolution of nonscalarized initial data indicated by black dots in Fig.~\ref{fig:initial_data}, for different values of the scalar field mass $\mu$. Figures~\ref{fig:time_evolution_xi15}~and~\ref{fig:time_evolution_xi30} show the time evolution of the central values $\phi_c$ (top panel) and $\rho_c$ (bottom panel), in the cases of $\xi=15$ and $\xi=30$, respectively. In every case, we see a dynamical transition from nonscalarized to scalarized, quasistatic configurations which are consistent with the expected end-states indicated with color marks in Fig.~\ref{fig:initial_data}. The values of $\phi_c$ and $\rho_c$ for the expected end-states are indicated by horizontal, dashed lines of the corresponding color, and their parameters are summarized in Table \ref{tab:end_state}.

\begin{table*}[]
  \centering
 \begin{tabular}{c |c |c |c |c |c |c |c |c |c |c } 
 \hline\hline
 \multicolumn{1}{c|}{} & \multicolumn{5}{c}{$\xi=15$} &  \multicolumn{5}{|c}{$\xi=30$} \\
 \hline
 Color in Fig.~\ref{fig:initial_data} & Red & Green & Blue & Orange & Brown & Red & Green & Blue & Orange & Brown \\
 \hline
 $\mu$ & 0.0 & 0.1 & 0.4 & 0.5 & 0.8 & 0.0 & 0.1 & 0.4 & 0.5 & 0.8 \\
 \hline
 $\rho_c/\rho_0$ & $8.4130$ & $8.8678$ & $9.1919$ & $8.9893$ & $8.9245$ & $8.2781$ & $8.6799$ & $9.2225$ & $9.2678$ & $8.9667$ \\
 \hline
 $M$ [$M_\odot$] & $1.43006$ & $1.43763$ & $1.44820$ & $1.44898$ & $1.44897$ & $1.42541$ & $1.43253$ & $1.44407$ & $1.44624$ & $1.44898$ \\
 \hline
 $R$ [km] & $10.94$ & $10.57$ & $10.19$ & $10.20$ & $10.21$ & $11.07$ & $10.74$ & $10.28$ & $10.21$ & $10.21$ \\
 \hline
 $|\phi_c|$ & $0.0199$ & $0.0173$ & $0.0092$ & $0.0040$ & $0.0$ & $0.0149$ & $0.0134$ & $0.0099$ & $0.0086$ & $0.0023$ \\
 \hline\hline
\end{tabular}
  \caption{Parameters of expected end-states of time evolution. They correspond to color marks in the bottom panel of Fig.~\ref{fig:initial_data}. All of the configurations have $M_b = 1.636 M_\odot$.}
  \label{tab:end_state}
\end{table*}

We find different, somewhat uneven, types of dynamical behavior, depending on the vale of $\mu$. Star oscillations are most strongly damped in the massless case, and the solution quickly settles around the expected scalarized configuration. This result is similar to those reported in Ref.~\cite{Degollado:2020lsa}. We have run a case of small field mass, namely $\mu=0.01$, and we find a dynamics very similar to the massless case---thus not shown---suggesting that the limit of $\mu=0$ is reached smoothly.
However, the qualitative picture changes as we look at the case of $\mu=0.1$, where damping is notably less aggressive, and long-lived oscillation modes get excited. A systematic study of oscillation spectra would be interesting, as well as comparison with predictions from linear perturbation theory, as reported in the massless case~\cite{Mendes:2018qwo}. Such tasks are beyond the goals of this work.


Unlike the cases of $\mu = \{0, 0.1\}$, the oscillation amplitude for $\mu > 0.1$ is relatively larger for $\phi_c$ than $\rho_c$. These oscillations are of higher frequency, and they damp at different rates depending on the vale of $\mu$. Regardless of the details of the transient phase, in all cases we find strong indications that, at late times, the solutions tend to stabilize around quasistatic states corresponding to the expected scalarized configurations. In the particular case of $\xi=15$ and $\mu=0.8$, we notice that the initial data remains unscalarized and basically unchanged, in the sense that $\phi_c$ keeps oscillating around 0 with a tiny amplitude of the order of 10$^{-10}$, and $\rho_c$ keeps oscillating around the initial central baryon-mass density. This behavior is also expected, and is consistent with the fact that there is no static, scalarized solution with baryon mass $M_b=1.636$ in this case---see Sec.~\ref{sec:static_solutions}.

\begin{figure}
    \centering
    \includegraphics[width=0.5\textwidth, angle=0]{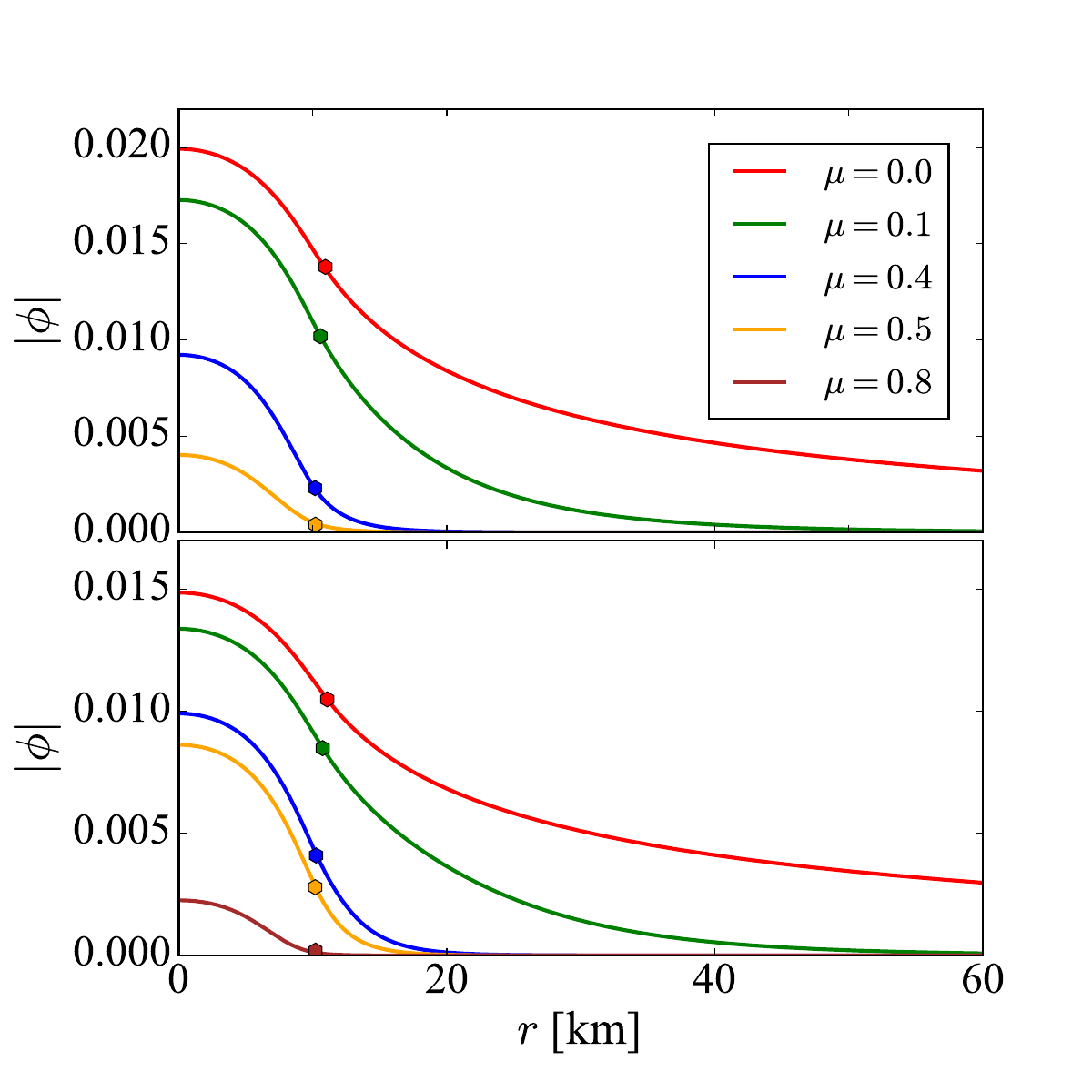}
    \caption{Scalar-field radial profiles of static configurations consistent with the final state of evolutions shown in Figs.~\ref{fig:time_evolution_xi15}~and~\ref{fig:time_evolution_xi30}. Top (bottom) panel corresponds to $\xi=15$ ($\xi=30$). Colored points indicate stellar radii. As the value of $\mu$ increases, the scalar field profiles get suppressed at shorter distances.}
    \label{fig:phi_profiles}
\end{figure}

An important effect of a massive scalar field during the scalarization process of a NS is that the amplitude and radial extent of the scalar field are suppressed as $\mu$ increases. This behavior can be appreciated from Fig.~\ref{fig:phi_profiles}, which shows scalar field profiles corresponding to the static solutions expected to be the end-state of time evolutions displayed in Figs.~\ref{fig:time_evolution_xi15}~and~\ref{fig:time_evolution_xi30}. Notice how the scalar-field cloud hardly extends beyond the stellar surface---indicated by colored marks---in the cases of $\xi=15$, $\mu=0.5$, and $\xi=30$, $\mu=0.8$.

\begin{figure}
    \centering
    \includegraphics[width=0.5\textwidth, angle=0]{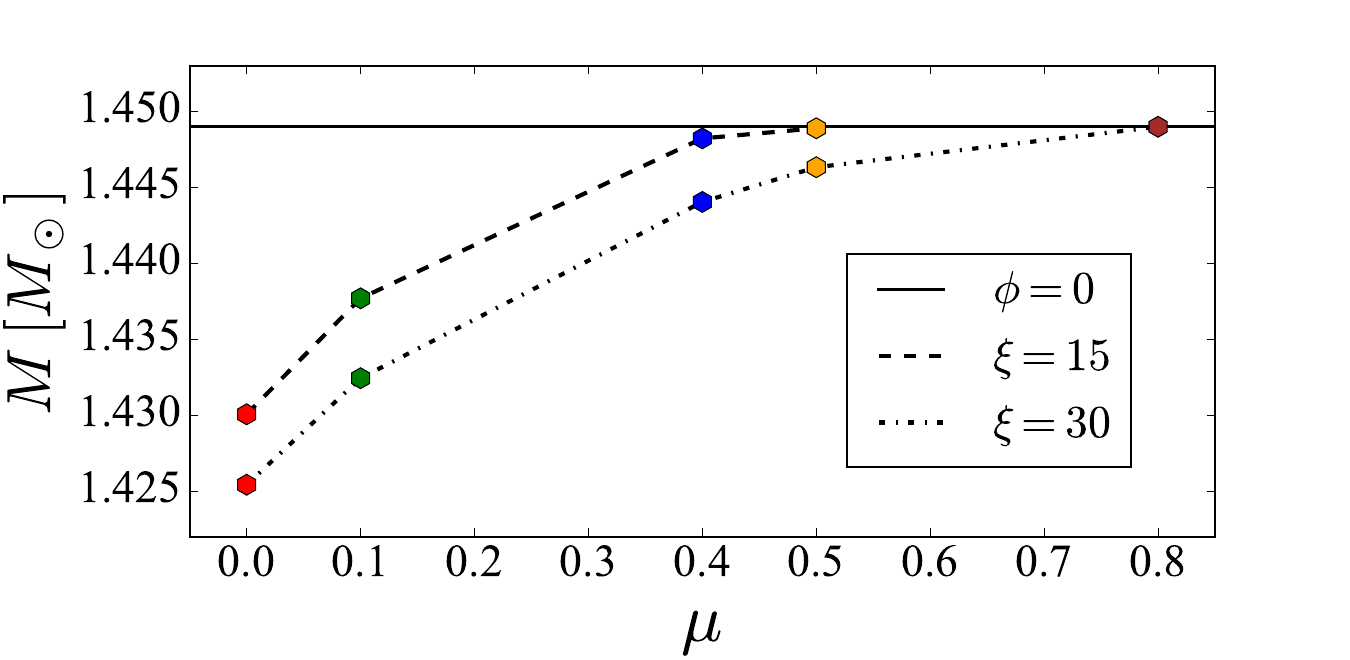}
\caption{Gravitational mass $M$ as a function of the scalar-field mass $\mu$, for the end-state of the dynamical processes of scalarization shown in Figs.~\ref{fig:time_evolution_xi15}~and~\ref{fig:time_evolution_xi30}. The solid, horizontal line indicates the corresponding mass of a (GR) unscalarized star.}
\label{fig:M_vs_m}
\end{figure}

Another relevant quantity of the static, scalarized solutions is their gravitational mass $M$, which is plotted as a function of $\mu$ in Fig.~\ref{fig:M_vs_m}, where the marks' color code is always consistent with previous figures. 
The smallest mass corresponds to the case of $\xi=30$ and $\mu=0$, which is the most scalarized configuration and therefore the one that differs the most from GR. As $\mu$ increases, all scalarized configurations tend to have the same gravitational mass as in GR, since the scalar field is highly suppressed despite the relatively high values of the NMC constant $\xi$.

\begin{figure}
    \centering
    \includegraphics[width=0.5\textwidth, angle=0]{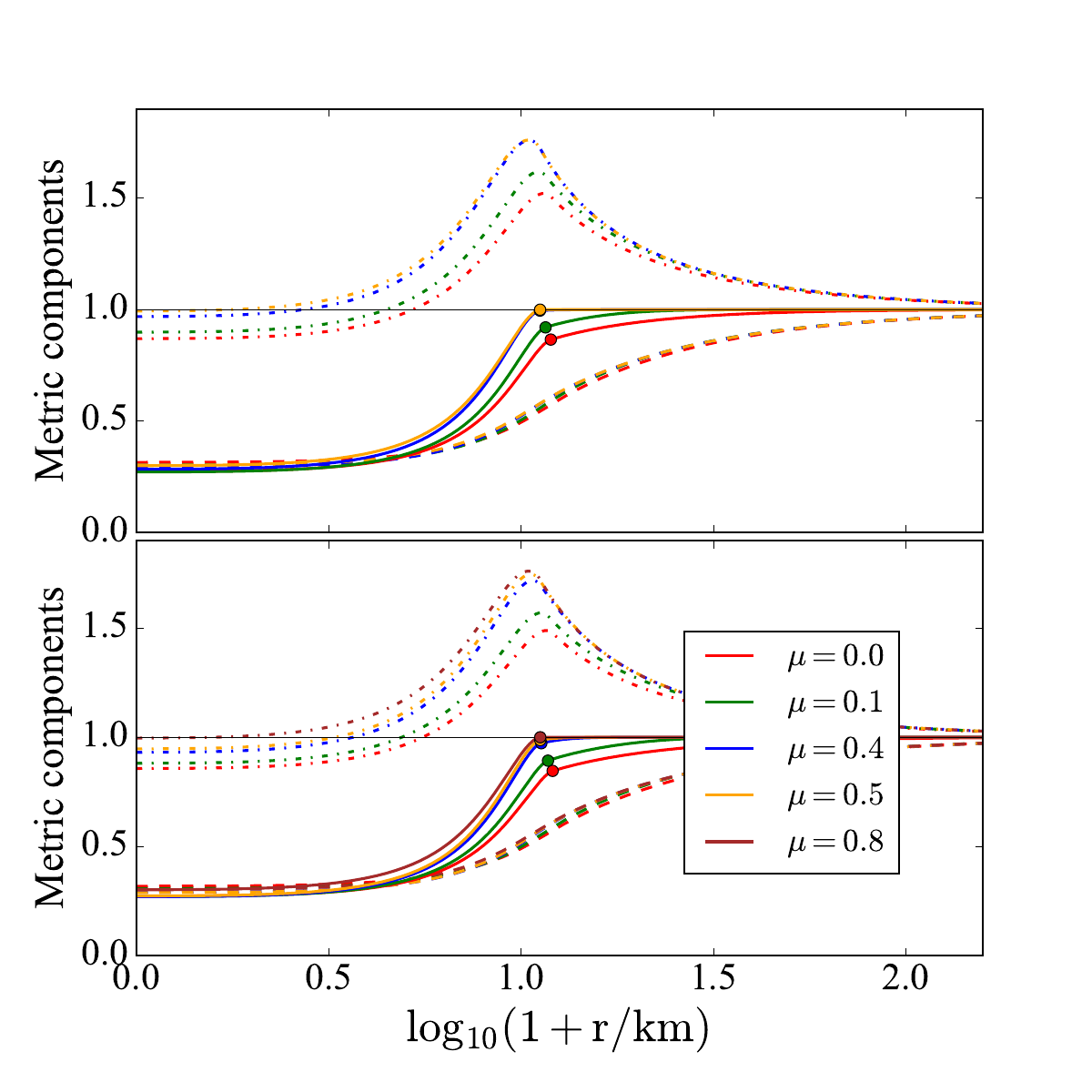}
\caption{Metric components of spacetimes consistent with the endpoint of evolutions shown in Figs.~\ref{fig:time_evolution_xi15}~and~\ref{fig:time_evolution_xi30}, in the cases of $\xi=15$ (top panel) and $\xi=30$ (bottom panel). Dashed lines correspond to $-g_{tt}$, dotted lines to $g_{rr}$, and solid lines to the product $-g_{tt} g_{rr}$. Circles indicate stellar surfaces.}
\label{fig:gttgrr}
\end{figure}

Although the spacetime of such spherical, static stars is asymptotically flat, it differs from the Schwarzschild solution outside the compact support of the fluid due to the scalar field contribution. This is manifested by the product of metric components $-g_{tt} g_{rr}$ (in area $r$-coordinates) which differs from one outside the star. This product is actually lower than one at the surface of stars, as shown in Fig.~\ref{fig:gttgrr}.\footnote{We construct stellar models in the EF such that $\tilde{g}_{rr} = 1$ in $r=0$. Transformation to the JF induces $g_{rr} < 1$ in $r=0$ for scalarized stars.}

\section{Discussion}
\label{sec:discussion}
Our results are consistent with those reported in Refs. \cite{Ramazanoglu:2016kul,Yazadjiev:2016pcb}, except that those authors do not study the dynamical transition to SC, but rather obtain the scalarized configurations by solving the field equations in a completely static \cite{Ramazanoglu:2016kul} or stationary (slow rotation) \cite{Yazadjiev:2016pcb} scenarios, and find the nontrivial scalar field configuration that vanishes asymptotically.
In principle, those configurations would correspond approximately to the ones reported here after evolution. But this is not exactly so, because the theory in Refs. \cite{Ramazanoglu:2016kul,Yazadjiev:2016pcb} is {\it not} the quadratic NMC STT that we use here. Furthermore, they perform the analysis in the EF with an exponential conformal factor which does not map to the JF theory used here. However, the most important point to stress is that, in principle, the suppression of the scalar field is such that one avoids the bounds imposed by the binary systems on the value of $\xi$, notably by PSR J0348+0432 \cite{Antoniadis2013} because the scalar field becomes short-ranged and would not have a noticeable effect on the companion, which orbits around $10^7$ km away from the scalarized NS, and also because possible GW-dipole radiation is also suppressed by the mass term \cite{Alsing2012}. This length scale of the binary system orbit translates into a mass $\mu \gg 10^{-16}$ eV \cite{Ramazanoglu:2016kul} in order to avoid the observational bounds.\footnote{In this section, we recover the units of $\mu$, as indicated in Sec.~\ref{sec:STT}.}
At the same time, one would expect the scalarization effects to be large enough so that SC leaves a trace that can be detected in some way, like by suitable detectors of a breathing mode of scalar waves during the scalarization transition, or as an imprint in the GW signal emitted, for instance, during binary NS mergers. This implies that the mass term should be small enough to avoid a total suppression of the scalarization effects inside NSs. If one takes the range of the scalar field to be larger than the NS radius $\sim 10$ km, this translates into $\mu\leq 10^{-9}$ eV \cite{Ramazanoglu:2016kul}. Thus, the range allowed for the mass becomes roughly $ 10^{-16} {\rm eV} \ll \mu \lesssim 10^{-9} {\rm eV}$. Finally, as argued in \cite{Ramazanoglu:2016kul}, the scalar-field mass can be further restricted due to high-spin superradiance effects in stellar-mass black holes \cite{Cardoso:2013fwa,Brito:2014wla,Brito:2015oca} to be outside the range $ 10^{-13} {\rm eV} \lesssim \mu \lesssim 10^{-11} {\rm eV}$. All these bounds lead us, for instance, to explore the mass $\mu$ in the domain $ 10^{-12} {\rm eV} \lesssim \mu \lesssim 10^{-9} {\rm eV}$ which is consistent with the range of masses specified in Sec.~\ref{sec:STT}.
In practice, this exploration shows that taking $\xi=15$ the scalar field is practically confined within the star for $10^{-9} {\rm eV} \lesssim \mu$, while for $\mu \lesssim 10^{-12} {\rm eV}$ the scalar field behaves as if it were massless. These remarks can be appreciated in Fig.~\ref{fig:phi_profiles}. 

Another important difference with respect to the analyses of \cite{Ramazanoglu:2016kul,Yazadjiev:2016pcb} is that their maximum mass configurations for scalarized static or stationary NSs are completely unavailable from initial data starting from unscalarized NSs (\textit{i.e.} NSs in GR) for a given EOS.
These remarks can be better appreciated from Fig.~\ref{fig:M_vs_m}, where the larger the NMC constant $\xi$ for a given mass $\mu$ and fixed EOS, the lower the gravitational mass, relative to GR, is found at the end of the dynamical scalarization process. This is at odds when one builds stationary NS configurations not from a dynamical process but starting from the structure equations in static situations, for instance. In that case, and as emphasized above, one can build scalarized NS configurations with a larger gravitational mass when increasing $\xi$ (cf. Ref.\cite{Salgado98}). In fact, by constructing static scalarized NSs in this way, and above the maximum baryon mass allowed in GR, one cannot even compare NS configurations using energetic arguments. In this sense, while maximum mass models of scalarized NSs above those of GR are interesting in many respects (they might explain the ``supermassive'' NSs observed recently with masses above $2M_\odot$ without the need of an \textit{ad hoc} EOS, see Ref. \cite{Degollado:2020lsa}) they are unavailable from a dynamical-process point of view, like the one analyzed here since, as we have stressed, {\it a priori} they cannot be obtained from GR initial data by definition. Again, this is because scalarized NSs always have a mass lower than the initial GR configuration that one started from after the dynamical transition, for a fixed total baryon mass.

A final technical point is in order. In the massless scalar field scenario one computes the {\it scalar charge} of the star (which, as we mentioned earlier, is the order parameter analog to magnetization in ferromagnets at low temperatures) by an integral that measures the {\it flux} of the gradient of the scalar field through a two-dimensional sphere at (spatial) infinity (cf. Eq.(7) of Ref.\cite{Degollado:2020lsa}). However, if one uses the same definition in the massive case, then the scalar charge vanishes due to the exponential fall-off of the field away from the star. This situation is similar to a scenario where the photon is massive, in which case, the electric charge of a point-source would vanish far away from the source due to the Yukawa fall-off of the electrostatic potential. Of course, operationally, one can still assign a value to the scalar charge as the coefficient of such a Yukawa term. However, by doing so, one would argue if such a value is a coordinate independent quantity, which does not seem evident {\it a priori}. Therefore, in the massive case it remains to analyze if the scalar charge thus ``defined'' operationally is indeed a good measure of the scalarization strength.

\section{Conclusions}
\label{sec:conclusions}
Following our previous numerical analysis about the dynamical transition to scalarized NSs in STT for a massless scalar field, and with a quadratic NMC to the Ricci curvature in the JF \cite{Degollado:2020lsa}, we perform here a similar analysis by adding a mass term in order to study its impact on the dynamics. Contrary to the massless scenario, the NMC constant $\xi$ can avoid the bounds imposed by several observations in binary systems since the mass term can suppress the range of the scalar field outside the star for a sufficiently large $\mu$, which is still far below the mass of any of the massive particles detected so far within the standard model of particle physics. 
Our results are consistent with previous studies \cite{Ramazanoglu:2016kul,Yazadjiev:2016pcb} where the scalarization process with a mass term was analyzed nondynamically, but differ drastically from them in that the initial available energy (the gravitational mass) leading to scalarized neutron stars is always, and by definition, larger than the gravitational mass of the final scalarized stationary star at the end of the process, where the energy difference is radiated away in the form of scalar GWs.
It is worth investigating if despite the scalar field suppression that allows the avoidance of available bounds with large values of the NMC constant $\xi$, the field still has some impact on other observational channels, like in GW signals from NS mergers observed by the present and future detectors, such as LIGO-VIRGO-KAGRA, the Einstein Telescope, and LISA.

\section*{Acknowledgments}
We acknowledge financial support through the Consejo Nacional de Humanidades, Ciencias y Tecnolog\'ias (CONAHCyT) grant ``Ciencia de Frontera" No. 140630, and Network Projects No. 376127 ``Sombras, lentes y ondas gravitatorias generadas por objetos compactos astrof\'\i sicos," and No. 304001 ``Estudio de campos escalares con aplicaciones en cosmolog\'ia y astrof\'isica." We also benefited from UNAM-PAPIIT grants No. IA101123, No. IN110523, and No. IN105223, and by the European Horizon Europe staff exchange (SE) programme HORIZON-MSCA2021-SE-01 Grant No. NewFun-FiCO101086251. Numerical simulations were performed at the \textsc{LAMOD} facility at ICN, UNAM.

\appendix

\section{Transformation between Einstein and Jordan frames}
\label{app:einstein_frame}
Consider the action of a STT formulated in the Jordan Frame~\cite{Damour96},\footnote{The authors of Ref.~\cite{Damour96} introduce a similar action but written in terms of a scalar field $\Phi$, related to $\phi$ in our Eq.~(\ref{eq:action}) by $\Phi= \sqrt{\kappa} \phi$. Thus the functions $F$, $Z$, and $U$ in~\cite{Damour96} do not coincide exactly with those in Eqns. (\ref{eq:F_app})-(\ref{eq:U_app}) for the specific STT studied here.}
\begin{eqnarray}
    S[g_{ab};\phi;\psi_\text{m}] &=&
    \frac{1}{2\kappa}\int \left[ F(\phi) R - Z(\phi) (\nabla\phi)^2 - U(\phi) \right] \sqrt{-g} d^4x \nonumber \\
    &+& S_\text{m}[g_{ab};\psi_\text{m}], \label{eq:S_JF}    
\end{eqnarray}
where $F$, $Z$, and $U$ are given functions of the scalar field $\phi$, $(\nabla\phi)^2 = g^{ab}\nabla_a\phi\nabla_b\phi$, $\kappa = 8\pi $, and $S_\text{m}$ is the action for matter fields $\psi_\text{m}$.

We assume a conformal rescaling of the metric, $g_{ab} = a(\tilde{\phi})^2 \tilde{g}_{ab}$, with a function $a(\tilde{\phi})$ of a scalar field $\tilde{\phi}$ related to $\phi$ through some function $\tilde{\phi}(\phi)$---see the details below.
The metric $\tilde{g}_{ab}$ is known as the \textit{Einstein-frame metric}. Under these assumptions, transforming the action~(\ref{eq:S_JF}) into the Einstein frame,
\begin{eqnarray}
    S[\tilde{g}_{ab};\tilde{\phi};\psi_\text{m}] &=& \frac{1}{2\kappa}\int \left[ \tilde{R} - 2 (\tilde{\nabla}\tilde{\phi})^2 - \tilde{V}(\tilde{\phi}) \right] \sqrt{-\tilde{g}} d^4x \nonumber \\
    &+& S_\text{m}[a(\tilde{\phi})^2\tilde{g}_{ab};\psi_\text{m}], \label{eq:S_EF}
\end{eqnarray}
demands
\begin{eqnarray}
a(\tilde{\phi})^2 &=& F(\phi(\tilde{\phi}))^{-1},\\
\tilde{V}(\tilde{\phi}) &=& F(\phi(\tilde{\phi}))^{-2} U(\phi(\tilde{\phi})),\\
\frac{d\tilde{\phi}}{d\phi} &=& \sqrt{\frac{3}{4}\left( \frac{F'(\phi)}{F(\phi)} \right)^2 + \frac{Z(\phi)}{2F(\phi)}}.
\end{eqnarray}

For the particular case of the action in Eq.~(\ref{eq:action}), we have
\begin{eqnarray}
F(\phi) &=& \kappa f(\phi), \label{eq:F_app}\\
Z(\phi) &=& \kappa, \label{eq:Z_app}\\
U(\phi) &=& 2\kappa V(\phi), \label{eq:U_app}
\end{eqnarray}
and then
\begin{eqnarray}
a(\tilde{\phi}) &=& \frac{1}{\sqrt{1 + \kappa\xi\phi(\tilde{\phi})^2}}, \label{eq:app_a} \\
\tilde{V}(\tilde{\phi}) &=& \frac{2\kappa}{\left[1+\kappa\xi\phi(\tilde{\phi})^2\right]^2} V(\phi(\tilde{\phi})), \label{eq:app_V}\\
\frac{d\tilde{\phi}}{d\phi} &=& \frac{\sqrt{\kappa/2}}{1+\kappa\xi\phi^2} \sqrt{1+\kappa\xi(1+6\xi)\phi^2}. \label{eq:app_phi}
\end{eqnarray}

We invert numerically the integral of Eq.~(\ref{eq:app_phi}), $\tilde{\phi}(\phi)$, in order to recover the scalar field in the Jordan Frame, $\phi(\tilde{\phi})$, which is required to evaluate the functions (\ref{eq:app_a}) and (\ref{eq:app_V}).

\section{Evolution equations}
\label{app:evolution_eqns}
The Einstein-scalar-field system resulting from variation of the \textit{Einstein frame} action (\ref{eq:S_EF}) is
\begin{eqnarray}
\tilde{G}_{ab} - 2 \tilde{\nabla}_a \tilde{\phi} \tilde{\nabla}_b\tilde{\phi} + 
\tilde{g}_{ab} \tilde{\nabla}_c \tilde{\phi} 
\tilde{\nabla}^c \tilde{\phi} + \frac{1}{2}\tilde{V}\tilde{g}_{ab} &=& 8\pi a^2 T_{ab}^\text{fluid},\label{eq:EF_field_eqns}\\
\tilde{\nabla}^b \tilde{\nabla}_b \tilde{\phi} - \frac{1}{4}\frac{d\tilde{V}}{d\tilde{\phi}} = -4\pi a^4 \alpha T_\text{fluid},&&\label{eq:EF_scalar_field_eqn}
\end{eqnarray}
where $\alpha = d \ln{a(\tilde{\phi})}/d\tilde{\phi}$, and $T_\text{fluid} = g_{ab}T^{ab}_\text{fluid}$.
We consider the spherically symmetric line element in Schwarzschild-like coordinates as
\begin{equation}
d\tilde{s}^2 = - \tilde{N}(t,r)^2 dt^2 + \tilde{A}(t,r)^2 dr^2 + r^2 \left( d\vartheta^2 + \sin^2\vartheta d\varphi^2\right),
\end{equation}
and we write the evolution equations from (\ref{eq:EF_field_eqns})--(\ref{eq:EF_scalar_field_eqn}) collectively with the hydrodynamic equations (\ref{eq:Tab_divergence})--(\ref{eq:continuity}) as a first-order system of conservation laws,
\begin{equation}\label{eq:flux-conservative-form}
\frac{\partial}{\partial t}(\tilde{A} {\bf q}) + \frac{1}{r^2} \frac{\partial}{\partial r}\left(\tilde{N}\tilde{A} r^2 {\bf F}({\bf q})\right) = {\bf S}({\bf q}),
\end{equation}
where ${\bf q} = ( D, S, \tau , \eta, \psi )^\text{T}$ includes conservative fluid quantities measured by Eulerian observers: baryon-mass density $D$, radial momentum density $S$, and internal energy-density $\tau$, defined by
\begin{eqnarray}
D &:=& \rho \Gamma,\\
S &:=& (E + p) A^2 v ,\\
\tau &:=& E - D,
\end{eqnarray}
where $\Gamma = \left(1 - \tilde{A}^2 v^2 \right)^{-1/2}$ is the Lorentz factor, $\tilde{A} v$ is the fluid's radial velocity, and $E = \Gamma^2 (\epsilon + p)\, -\, p$ is the total energy-density, all of them measured by Eulerian observers.
The scalar field variables in ${\bf q}$ are defined by
\begin{equation}
\eta := \frac{1}{\tilde{A}} \frac{\partial \tilde{\phi}}{\partial r}, \qquad
\psi := \frac{1}{\tilde{N}} \frac{\partial \tilde{\phi}}{\partial t}.
\end{equation}
The flux and source vectors in Eq.~(\ref{eq:flux-conservative-form}) are, respectively,
\begin{eqnarray}
{\bf F}({\bf q}) &=& (F_D, F_S, F_\tau, F_\eta, F_\psi)^\text{T}, \\
{\bf S}({\bf q}) &=& (S_D, S_S, S_\tau, S_\eta, S_\psi)^\text{T},
\end{eqnarray}
with
\begin{align}
F_D & = D v, \\
F_S & = S v + p, \\
F_\tau & = (\tau + p) v, \\ 
F_\eta & = -\psi/A, \\
F_\psi & = -\eta/A,
\end{align}
and
\begin{align}
S_D =& -3 \alpha \tilde{N} \tilde{A} D \left( \psi + \tilde{A}v \eta \right), \\
S_S =& 2 \tilde{N} \tilde{A} \frac{p}{r} 
	- \tilde{N}\tilde{A}^3 \frac{\tilde{m}}{r^2} (D + \tau + Sv + p) \nonumber \\
	& - 4 \alpha \tilde{N}\tilde{A} \psi S -\alpha \tilde{N}\tilde{A}^2 \eta (D + \tau + 3 S v + p) \nonumber \\
	& - \frac{1}{2} \tilde{N}\tilde{A}^3 r \left( \eta^2+\psi^2 \right) 
	(D + \tau - S v - p) \nonumber \\
        & + \frac{1}{4} \tilde{N}\tilde{A}^3 \tilde{V}r ( D + \tau + Sv + p),\\
S_\tau =& -\tilde{N}\tilde{A} S \frac{\tilde{m}}{r^2} - \alpha \tilde{N}\tilde{A} \psi 
	(3 \tau + S v + 3p) \nonumber \\
	& - \alpha \tilde{N}\tilde{A}^2 \eta v ( D + 4(\tau + p)) -\tilde{N}\tilde{A}^2 r \psi \eta (S v + p) \nonumber \\
	& - \frac{1}{2} \tilde{N}\tilde{A} S r \left(\eta^2 + \psi^2 - \tilde{V}/2\right), \\ 
S_\eta =& -2\tilde{N} \frac{\psi}{r}, \\
S_\psi =& - 4\pi \alpha a^4 \tilde{N}\tilde{A} \left(D + \tau - S v - 3p \right) - \frac{1}{4}\tilde{N}\tilde{A}\frac{d\tilde{V}}{d\tilde{\phi}},
\end{align}
where $\tilde{m}(t,r)$ is defined such that $\tilde{A}(t,r) = [1-2 \tilde{m}(t,r)/r ]^{-1/2}$. This mass aspect function evolves according to
\begin{equation}
\frac{\partial \tilde{m}}{\partial t} = r^2 \frac{\tilde{N}}{\tilde{A}^2} \left( \tilde{A} \eta\psi - 4\pi a^4 S \right).
\end{equation}
Moreover, at each time step, we enforce the lapse condition
\begin{equation}\label{eq:lapse_condition}
\frac{\partial \tilde{N}}{\partial r} = \tilde{A}^2 \tilde{N} \left[ \frac{\tilde{m}}{r^2} + 4\pi r a^4 (Sv + p) + \frac{r}{2} \left( \eta^2+\psi^2 - \tilde{V}/2 \right) \right].
\end{equation}
Finally, the baryonic mass is given by
\begin{equation}
M_b = \int_0^{R_s} 4 \pi r^2 D~a(\tilde{\phi})^3( 1- 2\tilde{m}/r)^{-1/2} dr.
\end{equation}

In the massless case, the evolution system above reduces to the one of Ref.~\cite{Mendes:2016fby}.

\bibliographystyle{apsrev4-1}
\nocite{apsrev41Control}
\bibliography{biboptions,referencias}

\begin{thebibliography}{60}%
\makeatletter
\providecommand \@ifxundefined [1]{%
 \@ifx{#1\undefined}
}%
\providecommand \@ifnum [1]{%
 \ifnum #1\expandafter \@firstoftwo
 \else \expandafter \@secondoftwo
 \fi
}%
\providecommand \@ifx [1]{%
 \ifx #1\expandafter \@firstoftwo
 \else \expandafter \@secondoftwo
 \fi
}%
\providecommand \natexlab [1]{#1}%
\providecommand \enquote  [1]{``#1''}%
\providecommand \bibnamefont  [1]{#1}%
\providecommand \bibfnamefont [1]{#1}%
\providecommand \citenamefont [1]{#1}%
\providecommand \href@noop [0]{\@secondoftwo}%
\providecommand \href [0]{\begingroup \@sanitize@url \@href}%
\providecommand \@href[1]{\@@startlink{#1}\@@href}%
\providecommand \@@href[1]{\endgroup#1\@@endlink}%
\providecommand \@sanitize@url [0]{\catcode `\\12\catcode `\$12\catcode
  `\&12\catcode `\#12\catcode `\^12\catcode `\_12\catcode `\%12\relax}%
\providecommand \@@startlink[1]{}%
\providecommand \@@endlink[0]{}%
\providecommand \url  [0]{\begingroup\@sanitize@url \@url }%
\providecommand \@url [1]{\endgroup\@href {#1}{\urlprefix }}%
\providecommand \urlprefix  [0]{URL }%
\providecommand \Eprint [0]{\href }%
\providecommand \doibase [0]{http://dx.doi.org/}%
\providecommand \selectlanguage [0]{\@gobble}%
\providecommand \bibinfo  [0]{\@secondoftwo}%
\providecommand \bibfield  [0]{\@secondoftwo}%
\providecommand \translation [1]{[#1]}%
\providecommand \BibitemOpen [0]{}%
\providecommand \bibitemStop [0]{}%
\providecommand \bibitemNoStop [0]{.\EOS\space}%
\providecommand \EOS [0]{\spacefactor3000\relax}%
\providecommand \BibitemShut  [1]{\csname bibitem#1\endcsname}%
\let\auto@bib@innerbib\@empty
\bibitem [{\citenamefont {Damour}\ and\ \citenamefont
  {Esposito-Farese}(1992)}]{Damour:1992we}%
  \BibitemOpen
  \bibfield  {author} {\bibinfo {author} {\bibfnamefont {T.}~\bibnamefont
  {Damour}}\ and\ \bibinfo {author} {\bibfnamefont {G.}~\bibnamefont
  {Esposito-Farese}},\ }\bibfield  {title} {\enquote {\bibinfo {title} {{Tensor
  multiscalar theories of gravitation}},}\ }\href {\doibase
  10.1088/0264-9381/9/9/015} {\bibfield  {journal} {\bibinfo  {journal} {Class.
  Quant. Grav.}\ }\textbf {\bibinfo {volume} {9}},\ \bibinfo {pages}
  {2093--2176} (\bibinfo {year} {1992})}\BibitemShut {NoStop}%
\bibitem [{\citenamefont {Fujii}\ and\ \citenamefont
  {Maeda}(2003)}]{Fujii2003}%
  \BibitemOpen
  \bibfield  {author} {\bibinfo {author} {\bibfnamefont {Y.}~\bibnamefont
  {Fujii}}\ and\ \bibinfo {author} {\bibfnamefont {K.-i.}\ \bibnamefont
  {Maeda}},\ }\href@noop {} {\emph {\bibinfo {title} {The Scalar-Tensor Theory
  of Gravitation}}},\ Cambridge Monographs on Mathematical Physics\ (\bibinfo
  {publisher} {Cambridge University Press},\ \bibinfo {year}
  {2003})\BibitemShut {NoStop}%
\bibitem [{\citenamefont {Brans}\ and\ \citenamefont
  {Dicke}(1961)}]{Brans:1961sx}%
  \BibitemOpen
  \bibfield  {author} {\bibinfo {author} {\bibfnamefont {C.}~\bibnamefont
  {Brans}}\ and\ \bibinfo {author} {\bibfnamefont {R.~H.}\ \bibnamefont
  {Dicke}},\ }\bibfield  {title} {\enquote {\bibinfo {title} {{Mach's principle
  and a relativistic theory of gravitation}},}\ }\href {\doibase
  10.1103/PhysRev.124.925} {\bibfield  {journal} {\bibinfo  {journal} {Phys.
  Rev.}\ }\textbf {\bibinfo {volume} {124}},\ \bibinfo {pages} {925--935}
  (\bibinfo {year} {1961})}\BibitemShut {NoStop}%
\bibitem [{\citenamefont {Brans}(1962)}]{Brans:1962zz}%
  \BibitemOpen
  \bibfield  {author} {\bibinfo {author} {\bibfnamefont {C.~H.}\ \bibnamefont
  {Brans}},\ }\bibfield  {title} {\enquote {\bibinfo {title} {{Mach's Principle
  and a Relativistic Theory of Gravitation. II}},}\ }\href {\doibase
  10.1103/PhysRev.125.2194} {\bibfield  {journal} {\bibinfo  {journal} {Phys.
  Rev.}\ }\textbf {\bibinfo {volume} {125}},\ \bibinfo {pages} {2194--2201}
  (\bibinfo {year} {1962})}\BibitemShut {NoStop}%
\bibitem [{\citenamefont {Dicke}(1962)}]{Dicke62}%
  \BibitemOpen
  \bibfield  {author} {\bibinfo {author} {\bibfnamefont {R.~H.}\ \bibnamefont
  {Dicke}},\ }\bibfield  {title} {\enquote {\bibinfo {title} {Mach's principle
  and invariance under transformation of units},}\ }\href {\doibase
  10.1103/PhysRev.125.2163} {\bibfield  {journal} {\bibinfo  {journal} {Phys.
  Rev.}\ }\textbf {\bibinfo {volume} {125}},\ \bibinfo {pages} {2163--2167}
  (\bibinfo {year} {1962})}\BibitemShut {NoStop}%
\bibitem [{\citenamefont {Broadhurst}\ \emph {et~al.}(1990)\citenamefont
  {Broadhurst}, \citenamefont {Ellis}, \citenamefont {Koo},\ and\ \citenamefont
  {Szalay}}]{Broadhurst:1990be}%
  \BibitemOpen
  \bibfield  {author} {\bibinfo {author} {\bibfnamefont {T.~J.}\ \bibnamefont
  {Broadhurst}}, \bibinfo {author} {\bibfnamefont {R.~S.}\ \bibnamefont
  {Ellis}}, \bibinfo {author} {\bibfnamefont {D.~C.}\ \bibnamefont {Koo}}, \
  and\ \bibinfo {author} {\bibfnamefont {A.~S.}\ \bibnamefont {Szalay}},\
  }\bibfield  {title} {\enquote {\bibinfo {title} {{Large scale distribution of
  galaxies at the galactic poles}},}\ }\href {\doibase 10.1038/343726a0}
  {\bibfield  {journal} {\bibinfo  {journal} {Nature}\ }\textbf {\bibinfo
  {volume} {343}},\ \bibinfo {pages} {726--728} (\bibinfo {year}
  {1990})}\BibitemShut {NoStop}%
\bibitem [{\citenamefont {{Morikawa}}(1990)}]{Morikawa90}%
  \BibitemOpen
  \bibfield  {author} {\bibinfo {author} {\bibfnamefont {M.}~\bibnamefont
  {{Morikawa}}},\ }\bibfield  {title} {\enquote {\bibinfo {title} {{Oscillating
  Universe: The Periodic Redshift Distribution of Galaxies with a Scale 128 H
  -1 Megaparsecs at the Galactic Poles}},}\ }\href {\doibase 10.1086/185842}
  {\bibfield  {journal} {\bibinfo  {journal} {\apjl}\ }\textbf {\bibinfo
  {volume} {362}},\ \bibinfo {pages} {L37} (\bibinfo {year}
  {1990})}\BibitemShut {NoStop}%
\bibitem [{\citenamefont {Salgado}\ \emph {et~al.}(1996)\citenamefont
  {Salgado}, \citenamefont {Sudarsky},\ and\ \citenamefont
  {Quevedo}}]{Salgado96}%
  \BibitemOpen
  \bibfield  {author} {\bibinfo {author} {\bibfnamefont {M.}~\bibnamefont
  {Salgado}}, \bibinfo {author} {\bibfnamefont {D.}~\bibnamefont {Sudarsky}}, \
  and\ \bibinfo {author} {\bibfnamefont {H.}~\bibnamefont {Quevedo}},\
  }\bibfield  {title} {\enquote {\bibinfo {title} {{Galactic periodicity and
  the oscillating G model}},}\ }\href {\doibase 10.1103/PhysRevD.53.6771}
  {\bibfield  {journal} {\bibinfo  {journal} {Phys. Rev. D}\ }\textbf {\bibinfo
  {volume} {53}},\ \bibinfo {pages} {6771--6783} (\bibinfo {year} {1996})},\
  \Eprint {http://arxiv.org/abs/gr-qc/9606038} {arXiv:gr-qc/9606038}
  \BibitemShut {NoStop}%
\bibitem [{\citenamefont {Salgado}\ \emph {et~al.}(1997)\citenamefont
  {Salgado}, \citenamefont {Sudarsky},\ and\ \citenamefont
  {Quevedo}}]{Salgado97a}%
  \BibitemOpen
  \bibfield  {author} {\bibinfo {author} {\bibfnamefont {M.}~\bibnamefont
  {Salgado}}, \bibinfo {author} {\bibfnamefont {D.}~\bibnamefont {Sudarsky}}, \
  and\ \bibinfo {author} {\bibfnamefont {H.}~\bibnamefont {Quevedo}},\
  }\bibfield  {title} {\enquote {\bibinfo {title} {{Has cosmological dark
  matter been observed?}}}\ }\href {\doibase 10.1016/S0370-2693(97)00798-3}
  {\bibfield  {journal} {\bibinfo  {journal} {Phys. Lett. B}\ }\textbf
  {\bibinfo {volume} {408}},\ \bibinfo {pages} {69--74} (\bibinfo {year}
  {1997})},\ \Eprint {http://arxiv.org/abs/gr-qc/9606039} {arXiv:gr-qc/9606039}
  \BibitemShut {NoStop}%
\bibitem [{\citenamefont {Quevedo}\ \emph {et~al.}(1997)\citenamefont
  {Quevedo}, \citenamefont {Salgado},\ and\ \citenamefont
  {Sudarsky}}]{Salgado97b}%
  \BibitemOpen
  \bibfield  {author} {\bibinfo {author} {\bibfnamefont {H.}~\bibnamefont
  {Quevedo}}, \bibinfo {author} {\bibfnamefont {M.}~\bibnamefont {Salgado}}, \
  and\ \bibinfo {author} {\bibfnamefont {D.}~\bibnamefont {Sudarsky}},\
  }\bibfield  {title} {\enquote {\bibinfo {title} {{The oscillating G model: A
  possible explanation for the nature of cosmological nonbaryonic matter}},}\
  }\href {\doibase 10.1086/304694} {\bibfield  {journal} {\bibinfo  {journal}
  {Astrophys. J.}\ }\textbf {\bibinfo {volume} {488}},\ \bibinfo {pages}
  {14--26} (\bibinfo {year} {1997})}\BibitemShut {NoStop}%
\bibitem [{\citenamefont {Boisseau}\ \emph {et~al.}(2000)\citenamefont
  {Boisseau}, \citenamefont {Esposito-Farese}, \citenamefont {Polarski},\ and\
  \citenamefont {Starobinsky}}]{Boisseau:2000pr}%
  \BibitemOpen
  \bibfield  {author} {\bibinfo {author} {\bibfnamefont {B.}~\bibnamefont
  {Boisseau}}, \bibinfo {author} {\bibfnamefont {G.}~\bibnamefont
  {Esposito-Farese}}, \bibinfo {author} {\bibfnamefont {D.}~\bibnamefont
  {Polarski}}, \ and\ \bibinfo {author} {\bibfnamefont {A.~A.}\ \bibnamefont
  {Starobinsky}},\ }\bibfield  {title} {\enquote {\bibinfo {title}
  {{Reconstruction of a scalar tensor theory of gravity in an accelerating
  universe}},}\ }\href {\doibase 10.1103/PhysRevLett.85.2236} {\bibfield
  {journal} {\bibinfo  {journal} {Phys. Rev. Lett.}\ }\textbf {\bibinfo
  {volume} {85}},\ \bibinfo {pages} {2236} (\bibinfo {year} {2000})},\ \Eprint
  {http://arxiv.org/abs/gr-qc/0001066} {arXiv:gr-qc/0001066} \BibitemShut
  {NoStop}%
\bibitem [{\citenamefont {Amendola}(2001)}]{Amendola2001}%
  \BibitemOpen
  \bibfield  {author} {\bibinfo {author} {\bibfnamefont {L.}~\bibnamefont
  {Amendola}},\ }\bibfield  {title} {\enquote {\bibinfo {title} {{Dark energy
  and the Boomerang data}},}\ }\href {\doibase 10.1103/PhysRevLett.86.196}
  {\bibfield  {journal} {\bibinfo  {journal} {Phys. Rev. Lett.}\ }\textbf
  {\bibinfo {volume} {86}},\ \bibinfo {pages} {196--199} (\bibinfo {year}
  {2001})},\ \Eprint {http://arxiv.org/abs/astro-ph/0006300}
  {arXiv:astro-ph/0006300} \BibitemShut {NoStop}%
\bibitem [{\citenamefont {Riazuelo}\ and\ \citenamefont
  {Uzan}(2002)}]{Riazuelo2002}%
  \BibitemOpen
  \bibfield  {author} {\bibinfo {author} {\bibfnamefont {A.}~\bibnamefont
  {Riazuelo}}\ and\ \bibinfo {author} {\bibfnamefont {J.-P.}\ \bibnamefont
  {Uzan}},\ }\bibfield  {title} {\enquote {\bibinfo {title} {{Cosmological
  observations in scalar - tensor quintessence}},}\ }\href {\doibase
  10.1103/PhysRevD.66.023525} {\bibfield  {journal} {\bibinfo  {journal} {Phys.
  Rev. D}\ }\textbf {\bibinfo {volume} {66}},\ \bibinfo {pages} {023525}
  (\bibinfo {year} {2002})},\ \Eprint {http://arxiv.org/abs/astro-ph/0107386}
  {arXiv:astro-ph/0107386} \BibitemShut {NoStop}%
\bibitem [{\citenamefont {Schimd}\ \emph {et~al.}(2005)\citenamefont {Schimd},
  \citenamefont {Uzan},\ and\ \citenamefont {Riazuelo}}]{Schimd2005}%
  \BibitemOpen
  \bibfield  {author} {\bibinfo {author} {\bibfnamefont {C.}~\bibnamefont
  {Schimd}}, \bibinfo {author} {\bibfnamefont {J.-P.}\ \bibnamefont {Uzan}}, \
  and\ \bibinfo {author} {\bibfnamefont {A.}~\bibnamefont {Riazuelo}},\
  }\bibfield  {title} {\enquote {\bibinfo {title} {{Weak lensing in
  scalar-tensor theories of gravity}},}\ }\href {\doibase
  10.1103/PhysRevD.71.083512} {\bibfield  {journal} {\bibinfo  {journal} {Phys.
  Rev. D}\ }\textbf {\bibinfo {volume} {71}},\ \bibinfo {pages} {083512}
  (\bibinfo {year} {2005})},\ \Eprint {http://arxiv.org/abs/astro-ph/0412120}
  {arXiv:astro-ph/0412120} \BibitemShut {NoStop}%
\bibitem [{\citenamefont {Flanagan}(2004)}]{Flanagan:2004bz}%
  \BibitemOpen
  \bibfield  {author} {\bibinfo {author} {\bibfnamefont {E.~E.}\ \bibnamefont
  {Flanagan}},\ }\bibfield  {title} {\enquote {\bibinfo {title} {{The Conformal
  frame freedom in theories of gravitation}},}\ }\href {\doibase
  10.1088/0264-9381/21/15/N02} {\bibfield  {journal} {\bibinfo  {journal}
  {Class. Quant. Grav.}\ }\textbf {\bibinfo {volume} {21}},\ \bibinfo {pages}
  {3817} (\bibinfo {year} {2004})},\ \Eprint
  {http://arxiv.org/abs/gr-qc/0403063} {arXiv:gr-qc/0403063} \BibitemShut
  {NoStop}%
\bibitem [{\citenamefont {Salgado}(2006)}]{Salgado06}%
  \BibitemOpen
  \bibfield  {author} {\bibinfo {author} {\bibfnamefont {M.}~\bibnamefont
  {Salgado}},\ }\bibfield  {title} {\enquote {\bibinfo {title} {{The Cauchy
  problem of scalar tensor theories of gravity}},}\ }\href {\doibase
  10.1088/0264-9381/23/14/010} {\bibfield  {journal} {\bibinfo  {journal}
  {Class. Quant. Grav.}\ }\textbf {\bibinfo {volume} {23}},\ \bibinfo {pages}
  {4719--4742} (\bibinfo {year} {2006})},\ \Eprint
  {http://arxiv.org/abs/gr-qc/0509001} {arXiv:gr-qc/0509001} \BibitemShut
  {NoStop}%
\bibitem [{\citenamefont {Salgado}\ \emph {et~al.}(2008)\citenamefont
  {Salgado}, \citenamefont {Martinez-del Rio}, \citenamefont {Alcubierre},\
  and\ \citenamefont {N\'u{\~n}ez}}]{Salgado08}%
  \BibitemOpen
  \bibfield  {author} {\bibinfo {author} {\bibfnamefont {M.}~\bibnamefont
  {Salgado}}, \bibinfo {author} {\bibfnamefont {D.}~\bibnamefont {Martinez-del
  Rio}}, \bibinfo {author} {\bibfnamefont {M.}~\bibnamefont {Alcubierre}}, \
  and\ \bibinfo {author} {\bibfnamefont {D.}~\bibnamefont {N\'u{\~n}ez}},\
  }\bibfield  {title} {\enquote {\bibinfo {title} {{Hyperbolicity of
  scalar-tensor theories of gravity}},}\ }\href {\doibase
  10.1103/PhysRevD.77.104010} {\bibfield  {journal} {\bibinfo  {journal} {Phys.
  Rev. D}\ }\textbf {\bibinfo {volume} {77}},\ \bibinfo {pages} {104010}
  (\bibinfo {year} {2008})},\ \Eprint {http://arxiv.org/abs/0801.2372}
  {arXiv:0801.2372 [gr-qc]} \BibitemShut {NoStop}%
\bibitem [{\citenamefont {Berti}\ \emph {et~al.}(2015)\citenamefont {Berti}
  \emph {et~al.}}]{Berti:2015itd}%
  \BibitemOpen
  \bibfield  {author} {\bibinfo {author} {\bibfnamefont {E.}~\bibnamefont
  {Berti}} \emph {et~al.},\ }\bibfield  {title} {\enquote {\bibinfo {title}
  {{Testing General Relativity with Present and Future Astrophysical
  Observations}},}\ }\href {\doibase 10.1088/0264-9381/32/24/243001} {\bibfield
   {journal} {\bibinfo  {journal} {Class. Quantum Grav.}\ }\textbf {\bibinfo
  {volume} {32}},\ \bibinfo {pages} {243001} (\bibinfo {year} {2015})},\
  \Eprint {http://arxiv.org/abs/1501.07274} {arXiv:1501.07274 [gr-qc]}
  \BibitemShut {NoStop}%
\bibitem [{\citenamefont {Damour}\ and\ \citenamefont
  {Esposito-Far\`ese}(1993)}]{Damour93}%
  \BibitemOpen
  \bibfield  {author} {\bibinfo {author} {\bibfnamefont {T.}~\bibnamefont
  {Damour}}\ and\ \bibinfo {author} {\bibfnamefont {G.}~\bibnamefont
  {Esposito-Far\`ese}},\ }\bibfield  {title} {\enquote {\bibinfo {title}
  {Nonperturbative strong-field effects in tensor-scalar theories of
  gravitation},}\ }\href {\doibase 10.1103/PhysRevLett.70.2220} {\bibfield
  {journal} {\bibinfo  {journal} {Phys. Rev. Lett.}\ }\textbf {\bibinfo
  {volume} {70}},\ \bibinfo {pages} {2220--2223} (\bibinfo {year}
  {1993})}\BibitemShut {NoStop}%
\bibitem [{\citenamefont {Damour}\ and\ \citenamefont
  {Esposito-Farese}(1996)}]{Damour96}%
  \BibitemOpen
  \bibfield  {author} {\bibinfo {author} {\bibfnamefont {T.}~\bibnamefont
  {Damour}}\ and\ \bibinfo {author} {\bibfnamefont {G.}~\bibnamefont
  {Esposito-Farese}},\ }\bibfield  {title} {\enquote {\bibinfo {title} {{Tensor
  - scalar gravity and binary pulsar experiments}},}\ }\href {\doibase
  10.1103/PhysRevD.54.1474} {\bibfield  {journal} {\bibinfo  {journal} {Phys.
  Rev. D}\ }\textbf {\bibinfo {volume} {54}},\ \bibinfo {pages} {1474--1491}
  (\bibinfo {year} {1996})},\ \Eprint {http://arxiv.org/abs/gr-qc/9602056}
  {arXiv:gr-qc/9602056} \BibitemShut {NoStop}%
\bibitem [{LVK()}]{LVK}%
  \BibitemOpen
  \href {https://pnp.ligo.org/ppcomm/Papers.html} {\bibinfo  {journal}
  {LIGO-Virgo-KAGRA Publications}\ }\BibitemShut {NoStop}%
\bibitem [{\citenamefont {Sotani}\ and\ \citenamefont
  {Kokkotas}(2005)}]{Sotani05}%
  \BibitemOpen
\bibfield  {journal} {  }\bibfield  {author} {\bibinfo {author} {\bibfnamefont
  {H.}~\bibnamefont {Sotani}}\ and\ \bibinfo {author} {\bibfnamefont {K.~D.}\
  \bibnamefont {Kokkotas}},\ }\bibfield  {title} {\enquote {\bibinfo {title}
  {Stellar oscillations in scalar-tensor theory of gravity},}\ }\href@noop {}
  {\bibfield  {journal} {\bibinfo  {journal} {Phys. Rev.}\ }\textbf {\bibinfo
  {volume} {D71}},\ \bibinfo {pages} {124038} (\bibinfo {year} {2005})},\
  \Eprint {http://arxiv.org/abs/gr-qc/0506060} {gr-qc/0506060} \BibitemShut
  {NoStop}%
\bibitem [{\citenamefont {Novak}(1998{\natexlab{a}})}]{Novak:1998rk}%
  \BibitemOpen
  \bibfield  {author} {\bibinfo {author} {\bibfnamefont {J.}~\bibnamefont
  {Novak}},\ }\bibfield  {title} {\enquote {\bibinfo {title} {{Neutron star
  transition to strong scalar field state in tensor scalar gravity}},}\ }\href
  {\doibase 10.1103/PhysRevD.58.064019} {\bibfield  {journal} {\bibinfo
  {journal} {Phys. Rev. D}\ }\textbf {\bibinfo {volume} {58}},\ \bibinfo
  {pages} {064019} (\bibinfo {year} {1998}{\natexlab{a}})},\ \Eprint
  {http://arxiv.org/abs/gr-qc/9806022} {arXiv:gr-qc/9806022} \BibitemShut
  {NoStop}%
\bibitem [{\citenamefont {Novak}(1998{\natexlab{b}})}]{Novak98a}%
  \BibitemOpen
  \bibfield  {author} {\bibinfo {author} {\bibfnamefont {J.}~\bibnamefont
  {Novak}},\ }\bibfield  {title} {\enquote {\bibinfo {title} {Spherical neutron
  star collapse toward a black hole in a tensor-scalar theory of gravity},}\
  }\href {\doibase 10.1103/PhysRevD.57.4789} {\bibfield  {journal} {\bibinfo
  {journal} {Phys. Rev. D}\ }\textbf {\bibinfo {volume} {57}},\ \bibinfo
  {pages} {4789--4801} (\bibinfo {year} {1998}{\natexlab{b}})}\BibitemShut
  {NoStop}%
\bibitem [{\citenamefont {Whinnett}(2000)}]{Whinnett00}%
  \BibitemOpen
  \bibfield  {author} {\bibinfo {author} {\bibfnamefont {A.~W.}\ \bibnamefont
  {Whinnett}},\ }\bibfield  {title} {\enquote {\bibinfo {title} {{Spontaneous
  scalarization and boson stars}},}\ }\href {\doibase
  10.1103/PhysRevD.61.124014} {\bibfield  {journal} {\bibinfo  {journal} {Phys.
  Rev. D}\ }\textbf {\bibinfo {volume} {61}},\ \bibinfo {pages} {124014}
  (\bibinfo {year} {2000})},\ \Eprint {http://arxiv.org/abs/gr-qc/9911052}
  {arXiv:gr-qc/9911052} \BibitemShut {NoStop}%
\bibitem [{\citenamefont {Alcubierre}\ \emph {et~al.}(2010)\citenamefont
  {Alcubierre}, \citenamefont {Degollado}, \citenamefont {N\'u{\~n}ez},
  \citenamefont {Ruiz},\ and\ \citenamefont {Salgado}}]{Alcubierre:2010ea}%
  \BibitemOpen
  \bibfield  {author} {\bibinfo {author} {\bibfnamefont {M.}~\bibnamefont
  {Alcubierre}}, \bibinfo {author} {\bibfnamefont {J.~C.}\ \bibnamefont
  {Degollado}}, \bibinfo {author} {\bibfnamefont {D.}~\bibnamefont
  {N\'u{\~n}ez}}, \bibinfo {author} {\bibfnamefont {M.}~\bibnamefont {Ruiz}}, \
  and\ \bibinfo {author} {\bibfnamefont {M.}~\bibnamefont {Salgado}},\
  }\bibfield  {title} {\enquote {\bibinfo {title} {{Dynamic transition to
  spontaneous scalarization in boson stars}},}\ }\href {\doibase
  10.1103/PhysRevD.81.124018} {\bibfield  {journal} {\bibinfo  {journal} {Phys.
  Rev.}\ }\textbf {\bibinfo {volume} {D81}},\ \bibinfo {pages} {124018}
  (\bibinfo {year} {2010})},\ \Eprint {http://arxiv.org/abs/1003.4767}
  {arXiv:1003.4767 [gr-qc]} \BibitemShut {NoStop}%
\bibitem [{\citenamefont {Silva}\ \emph {et~al.}(2018)\citenamefont {Silva},
  \citenamefont {Sakstein}, \citenamefont {Gualtieri}, \citenamefont
  {Sotiriou},\ and\ \citenamefont {Berti}}]{Silva:2017uqg}%
  \BibitemOpen
  \bibfield  {author} {\bibinfo {author} {\bibfnamefont {H.~O.}\ \bibnamefont
  {Silva}}, \bibinfo {author} {\bibfnamefont {J.}~\bibnamefont {Sakstein}},
  \bibinfo {author} {\bibfnamefont {L.}~\bibnamefont {Gualtieri}}, \bibinfo
  {author} {\bibfnamefont {T.~P.}\ \bibnamefont {Sotiriou}}, \ and\ \bibinfo
  {author} {\bibfnamefont {E.}~\bibnamefont {Berti}},\ }\bibfield  {title}
  {\enquote {\bibinfo {title} {{Spontaneous scalarization of black holes and
  compact stars from a Gauss-Bonnet coupling}},}\ }\href {\doibase
  10.1103/PhysRevLett.120.131104} {\bibfield  {journal} {\bibinfo  {journal}
  {Phys. Rev. Lett.}\ }\textbf {\bibinfo {volume} {120}},\ \bibinfo {pages}
  {131104} (\bibinfo {year} {2018})},\ \Eprint
  {http://arxiv.org/abs/1711.02080} {arXiv:1711.02080 [gr-qc]} \BibitemShut
  {NoStop}%
\bibitem [{\citenamefont {Doneva}\ and\ \citenamefont
  {Yazadjiev}(2018)}]{Doneva:2017bvd}%
  \BibitemOpen
  \bibfield  {author} {\bibinfo {author} {\bibfnamefont {D.~D.}\ \bibnamefont
  {Doneva}}\ and\ \bibinfo {author} {\bibfnamefont {S.~S.}\ \bibnamefont
  {Yazadjiev}},\ }\bibfield  {title} {\enquote {\bibinfo {title} {{New
  Gauss-Bonnet Black Holes with Curvature-Induced Scalarization in Extended
  Scalar-Tensor Theories}},}\ }\href {\doibase 10.1103/PhysRevLett.120.131103}
  {\bibfield  {journal} {\bibinfo  {journal} {Phys. Rev. Lett.}\ }\textbf
  {\bibinfo {volume} {120}},\ \bibinfo {pages} {131103} (\bibinfo {year}
  {2018})},\ \Eprint {http://arxiv.org/abs/1711.01187} {arXiv:1711.01187
  [gr-qc]} \BibitemShut {NoStop}%
\bibitem [{\citenamefont {Doneva}\ \emph {et~al.}(2024)\citenamefont {Doneva},
  \citenamefont {Ramazano\u{g}lu}, \citenamefont {Silva}, \citenamefont
  {Sotiriou},\ and\ \citenamefont {Yazadjiev}}]{Doneva:2022ewd}%
  \BibitemOpen
  \bibfield  {author} {\bibinfo {author} {\bibfnamefont {D.~D.}\ \bibnamefont
  {Doneva}}, \bibinfo {author} {\bibfnamefont {F.~M.}\ \bibnamefont
  {Ramazano\u{g}lu}}, \bibinfo {author} {\bibfnamefont {H.~O.}\ \bibnamefont
  {Silva}}, \bibinfo {author} {\bibfnamefont {T.~P.}\ \bibnamefont {Sotiriou}},
  \ and\ \bibinfo {author} {\bibfnamefont {S.~S.}\ \bibnamefont {Yazadjiev}},\
  }\bibfield  {title} {\enquote {\bibinfo {title} {{Spontaneous
  scalarization}},}\ }\href {\doibase 10.1103/RevModPhys.96.015004} {\bibfield
  {journal} {\bibinfo  {journal} {Rev. Mod. Phys.}\ }\textbf {\bibinfo {volume}
  {96}},\ \bibinfo {pages} {015004} (\bibinfo {year} {2024})},\ \Eprint
  {http://arxiv.org/abs/2211.01766} {arXiv:2211.01766 [gr-qc]} \BibitemShut
  {NoStop}%
\bibitem [{\citenamefont {Salgado}\ \emph {et~al.}(1998)\citenamefont
  {Salgado}, \citenamefont {Sudarsky},\ and\ \citenamefont
  {Nucamendi}}]{Salgado98}%
  \BibitemOpen
  \bibfield  {author} {\bibinfo {author} {\bibfnamefont {M.}~\bibnamefont
  {Salgado}}, \bibinfo {author} {\bibfnamefont {D.}~\bibnamefont {Sudarsky}}, \
  and\ \bibinfo {author} {\bibfnamefont {U.}~\bibnamefont {Nucamendi}},\
  }\bibfield  {title} {\enquote {\bibinfo {title} {Spontaneous
  scalarization},}\ }\href {\doibase 10.1103/PhysRevD.58.124003} {\bibfield
  {journal} {\bibinfo  {journal} {Phys. Rev. D}\ }\textbf {\bibinfo {volume}
  {58}},\ \bibinfo {pages} {124003} (\bibinfo {year} {1998})}\BibitemShut
  {NoStop}%
\bibitem [{\citenamefont {Antoniadis}\ \emph {et~al.}(2013)\citenamefont
  {Antoniadis} \emph {et~al.}}]{Antoniadis2013}%
  \BibitemOpen
  \bibfield  {author} {\bibinfo {author} {\bibfnamefont {J.}~\bibnamefont
  {Antoniadis}} \emph {et~al.},\ }\bibfield  {title} {\enquote {\bibinfo
  {title} {{A Massive Pulsar in a Compact Relativistic Binary}},}\ }\href
  {\doibase 10.1126/science.1233232} {\bibfield  {journal} {\bibinfo  {journal}
  {Science}\ }\textbf {\bibinfo {volume} {340}},\ \bibinfo {pages} {6131}
  (\bibinfo {year} {2013})},\ \Eprint {http://arxiv.org/abs/1304.6875}
  {arXiv:1304.6875 [astro-ph.HE]} \BibitemShut {NoStop}%
\bibitem [{\citenamefont {Demorest}\ \emph {et~al.}(2010)\citenamefont
  {Demorest}, \citenamefont {Pennucci}, \citenamefont {Ransom}, \citenamefont
  {Roberts},\ and\ \citenamefont {Hessels}}]{Demorest2010}%
  \BibitemOpen
  \bibfield  {author} {\bibinfo {author} {\bibfnamefont {P.}~\bibnamefont
  {Demorest}}, \bibinfo {author} {\bibfnamefont {T.}~\bibnamefont {Pennucci}},
  \bibinfo {author} {\bibfnamefont {S.}~\bibnamefont {Ransom}}, \bibinfo
  {author} {\bibfnamefont {M.}~\bibnamefont {Roberts}}, \ and\ \bibinfo
  {author} {\bibfnamefont {J.}~\bibnamefont {Hessels}},\ }\bibfield  {title}
  {\enquote {\bibinfo {title} {{A two-solar-mass neutron star measured using
  Shapiro delay}},}\ }\href {\doibase 10.1038/nature09466} {\bibfield
  {journal} {\bibinfo  {journal} {Nature}\ }\textbf {\bibinfo {volume} {467}},\
  \bibinfo {pages} {1081--1083} (\bibinfo {year} {2010})},\ \Eprint
  {http://arxiv.org/abs/1010.5788} {arXiv:1010.5788 [astro-ph.HE]} \BibitemShut
  {NoStop}%
\bibitem [{\citenamefont {Cromartie}\ \emph {et~al.}(2019)\citenamefont
  {Cromartie} \emph {et~al.}}]{Cromartie2019}%
  \BibitemOpen
  \bibfield  {author} {\bibinfo {author} {\bibfnamefont {H.~T.}\ \bibnamefont
  {Cromartie}} \emph {et~al.} (\bibinfo {collaboration} {NANOGrav}),\
  }\bibfield  {title} {\enquote {\bibinfo {title} {{Relativistic Shapiro delay
  measurements of an extremely massive millisecond pulsar}},}\ }\href {\doibase
  10.1038/s41550-019-0880-2} {\bibfield  {journal} {\bibinfo  {journal} {Nature
  Astron.}\ }\textbf {\bibinfo {volume} {4}},\ \bibinfo {pages} {72--76}
  (\bibinfo {year} {2019})},\ \Eprint {http://arxiv.org/abs/1904.06759}
  {arXiv:1904.06759 [astro-ph.HE]} \BibitemShut {NoStop}%
\bibitem [{\citenamefont {Romani}\ \emph {et~al.}(2022)\citenamefont {Romani},
  \citenamefont {Kandel}, \citenamefont {Filippenko}, \citenamefont {Brink},\
  and\ \citenamefont {Zheng}}]{Romani2022}%
  \BibitemOpen
  \bibfield  {author} {\bibinfo {author} {\bibfnamefont {R.~W.}\ \bibnamefont
  {Romani}}, \bibinfo {author} {\bibfnamefont {D.}~\bibnamefont {Kandel}},
  \bibinfo {author} {\bibfnamefont {A.~V.}\ \bibnamefont {Filippenko}},
  \bibinfo {author} {\bibfnamefont {T.~G.}\ \bibnamefont {Brink}}, \ and\
  \bibinfo {author} {\bibfnamefont {W.}~\bibnamefont {Zheng}},\ }\bibfield
  {title} {\enquote {\bibinfo {title} {{PSR J0952\ensuremath{-}0607: The
  Fastest and Heaviest Known Galactic Neutron Star}},}\ }\href {\doibase
  10.3847/2041-8213/ac8007} {\bibfield  {journal} {\bibinfo  {journal}
  {Astrophys. J. Lett.}\ }\textbf {\bibinfo {volume} {934}},\ \bibinfo {pages}
  {L17} (\bibinfo {year} {2022})},\ \Eprint {http://arxiv.org/abs/2207.05124}
  {arXiv:2207.05124 [astro-ph.HE]} \BibitemShut {NoStop}%
\bibitem [{\citenamefont {Freire}\ \emph {et~al.}(2012)\citenamefont {Freire}
  \emph {et~al.}}]{Freire2012}%
  \BibitemOpen
  \bibfield  {author} {\bibinfo {author} {\bibfnamefont {P.~C.~C.}\
  \bibnamefont {Freire}} \emph {et~al.},\ }\bibfield  {title} {\enquote
  {\bibinfo {title} {{The relativistic pulsar-white dwarf binary {PSR}
  {J}1738+0333 - {II}. {T}he most stringent test of scalar-tensor gravity}},}\
  }\href {\doibase 10.1111/j.1365-2966.2012.21253.x} {\bibfield  {journal}
  {\bibinfo  {journal} {Mon. Not. Roy. Astron. Soc.}\ }\textbf {\bibinfo
  {volume} {423}},\ \bibinfo {pages} {3328} (\bibinfo {year} {2012})},\ \Eprint
  {http://arxiv.org/abs/1205.1450} {arXiv:1205.1450 [astro-ph.GA]} \BibitemShut
  {NoStop}%
\bibitem [{\citenamefont {Shao}\ \emph {et~al.}(2017)\citenamefont {Shao},
  \citenamefont {Sennett}, \citenamefont {Buonanno}, \citenamefont {Kramer},\
  and\ \citenamefont {Wex}}]{Shao:2017gwu}%
  \BibitemOpen
  \bibfield  {author} {\bibinfo {author} {\bibfnamefont {L.}~\bibnamefont
  {Shao}}, \bibinfo {author} {\bibfnamefont {N.}~\bibnamefont {Sennett}},
  \bibinfo {author} {\bibfnamefont {A.}~\bibnamefont {Buonanno}}, \bibinfo
  {author} {\bibfnamefont {M.}~\bibnamefont {Kramer}}, \ and\ \bibinfo {author}
  {\bibfnamefont {N.}~\bibnamefont {Wex}},\ }\bibfield  {title} {\enquote
  {\bibinfo {title} {{Constraining nonperturbative strong-field effects in
  scalar-tensor gravity by combining pulsar timing and laser-interferometer
  gravitational-wave detectors}},}\ }\href {\doibase 10.1103/PhysRevX.7.041025}
  {\bibfield  {journal} {\bibinfo  {journal} {Phys. Rev. X}\ }\textbf {\bibinfo
  {volume} {7}},\ \bibinfo {pages} {041025} (\bibinfo {year} {2017})},\ \Eprint
  {http://arxiv.org/abs/1704.07561} {arXiv:1704.07561 [gr-qc]} \BibitemShut
  {NoStop}%
\bibitem [{\citenamefont {Kramer}\ \emph {et~al.}(2021)\citenamefont {Kramer}
  \emph {et~al.}}]{Kramer:2021jcw}%
  \BibitemOpen
  \bibfield  {author} {\bibinfo {author} {\bibfnamefont {M.}~\bibnamefont
  {Kramer}} \emph {et~al.},\ }\bibfield  {title} {\enquote {\bibinfo {title}
  {{Strong-Field Gravity Tests with the Double Pulsar}},}\ }\href {\doibase
  10.1103/PhysRevX.11.041050} {\bibfield  {journal} {\bibinfo  {journal} {Phys.
  Rev. X}\ }\textbf {\bibinfo {volume} {11}},\ \bibinfo {pages} {041050}
  (\bibinfo {year} {2021})},\ \Eprint {http://arxiv.org/abs/2112.06795}
  {arXiv:2112.06795 [astro-ph.HE]} \BibitemShut {NoStop}%
\bibitem [{\citenamefont {Zhao}\ \emph {et~al.}(2022)\citenamefont {Zhao},
  \citenamefont {Freire}, \citenamefont {Kramer}, \citenamefont {Shao},\ and\
  \citenamefont {Wex}}]{Zhao:2022vig}%
  \BibitemOpen
  \bibfield  {author} {\bibinfo {author} {\bibfnamefont {J.}~\bibnamefont
  {Zhao}}, \bibinfo {author} {\bibfnamefont {P.~C.~C.}\ \bibnamefont {Freire}},
  \bibinfo {author} {\bibfnamefont {M.}~\bibnamefont {Kramer}}, \bibinfo
  {author} {\bibfnamefont {L.}~\bibnamefont {Shao}}, \ and\ \bibinfo {author}
  {\bibfnamefont {N.}~\bibnamefont {Wex}},\ }\bibfield  {title} {\enquote
  {\bibinfo {title} {{Closing a spontaneous-scalarization window with binary
  pulsars}},}\ }\href {\doibase 10.1088/1361-6382/ac69a3} {\bibfield  {journal}
  {\bibinfo  {journal} {Class. Quant. Grav.}\ }\textbf {\bibinfo {volume}
  {39}},\ \bibinfo {pages} {11LT01} (\bibinfo {year} {2022})},\ \Eprint
  {http://arxiv.org/abs/2201.03771} {arXiv:2201.03771 [astro-ph.HE]}
  \BibitemShut {NoStop}%
\bibitem [{\citenamefont {Mendes}\ and\ \citenamefont
  {Ottoni}(2019)}]{Mendes:2019zpw}%
  \BibitemOpen
  \bibfield  {author} {\bibinfo {author} {\bibfnamefont {R.~F.~P.}\
  \bibnamefont {Mendes}}\ and\ \bibinfo {author} {\bibfnamefont
  {T.}~\bibnamefont {Ottoni}},\ }\bibfield  {title} {\enquote {\bibinfo {title}
  {{Scalar charges and pulsar-timing observables in the presence of
  nonminimally coupled scalar fields}},}\ }\href {\doibase
  10.1103/PhysRevD.99.124003} {\bibfield  {journal} {\bibinfo  {journal} {Phys.
  Rev. D}\ }\textbf {\bibinfo {volume} {99}},\ \bibinfo {pages} {124003}
  (\bibinfo {year} {2019})},\ \Eprint {http://arxiv.org/abs/1903.11638}
  {arXiv:1903.11638 [gr-qc]} \BibitemShut {NoStop}%
\bibitem [{\citenamefont {Chen}\ \emph {et~al.}(2015)\citenamefont {Chen},
  \citenamefont {Suyama},\ and\ \citenamefont {Yokoyama}}]{PhysRevD.92.124016}%
  \BibitemOpen
  \bibfield  {author} {\bibinfo {author} {\bibfnamefont {P.}~\bibnamefont
  {Chen}}, \bibinfo {author} {\bibfnamefont {T.}~\bibnamefont {Suyama}}, \ and\
  \bibinfo {author} {\bibfnamefont {J.}~\bibnamefont {Yokoyama}},\ }\bibfield
  {title} {\enquote {\bibinfo {title} {Spontaneous-scalarization-induced dark
  matter and variation of the gravitational constant},}\ }\href {\doibase
  10.1103/PhysRevD.92.124016} {\bibfield  {journal} {\bibinfo  {journal} {Phys.
  Rev. D}\ }\textbf {\bibinfo {volume} {92}},\ \bibinfo {pages} {124016}
  (\bibinfo {year} {2015})}\BibitemShut {NoStop}%
\bibitem [{\citenamefont {Ramazano\u{g}lu}\ and\ \citenamefont
  {Pretorius}(2016)}]{Ramazanoglu:2016kul}%
  \BibitemOpen
  \bibfield  {author} {\bibinfo {author} {\bibfnamefont {F.~M.}\ \bibnamefont
  {Ramazano\u{g}lu}}\ and\ \bibinfo {author} {\bibfnamefont {F.}~\bibnamefont
  {Pretorius}},\ }\bibfield  {title} {\enquote {\bibinfo {title} {{Spontaneous
  Scalarization with Massive Fields}},}\ }\href {\doibase
  10.1103/PhysRevD.93.064005} {\bibfield  {journal} {\bibinfo  {journal} {Phys.
  Rev. D}\ }\textbf {\bibinfo {volume} {93}},\ \bibinfo {pages} {064005}
  (\bibinfo {year} {2016})},\ \Eprint {http://arxiv.org/abs/1601.07475}
  {arXiv:1601.07475 [gr-qc]} \BibitemShut {NoStop}%
\bibitem [{\citenamefont {Yazadjiev}\ \emph {et~al.}(2016)\citenamefont
  {Yazadjiev}, \citenamefont {Doneva},\ and\ \citenamefont
  {Popchev}}]{Yazadjiev:2016pcb}%
  \BibitemOpen
  \bibfield  {author} {\bibinfo {author} {\bibfnamefont {S.~S.}\ \bibnamefont
  {Yazadjiev}}, \bibinfo {author} {\bibfnamefont {D.~D.}\ \bibnamefont
  {Doneva}}, \ and\ \bibinfo {author} {\bibfnamefont {D.}~\bibnamefont
  {Popchev}},\ }\bibfield  {title} {\enquote {\bibinfo {title} {{Slowly
  rotating neutron stars in scalar-tensor theories with a massive scalar
  field}},}\ }\href {\doibase 10.1103/PhysRevD.93.084038} {\bibfield  {journal}
  {\bibinfo  {journal} {Phys. Rev. D}\ }\textbf {\bibinfo {volume} {93}},\
  \bibinfo {pages} {084038} (\bibinfo {year} {2016})},\ \Eprint
  {http://arxiv.org/abs/1602.04766} {arXiv:1602.04766 [gr-qc]} \BibitemShut
  {NoStop}%
\bibitem [{\citenamefont {Doneva}\ and\ \citenamefont
  {Yazadjiev}(2016)}]{Doneva:2016xmf}%
  \BibitemOpen
  \bibfield  {author} {\bibinfo {author} {\bibfnamefont {D.~D.}\ \bibnamefont
  {Doneva}}\ and\ \bibinfo {author} {\bibfnamefont {S.~S.}\ \bibnamefont
  {Yazadjiev}},\ }\bibfield  {title} {\enquote {\bibinfo {title} {{Rapidly
  rotating neutron stars with a massive scalar field\textemdash{}structure and
  universal relations}},}\ }\href {\doibase 10.1088/1475-7516/2016/11/019}
  {\bibfield  {journal} {\bibinfo  {journal} {JCAP}\ }\textbf {\bibinfo
  {volume} {11}},\ \bibinfo {pages} {019} (\bibinfo {year} {2016})},\ \Eprint
  {http://arxiv.org/abs/1607.03299} {arXiv:1607.03299 [gr-qc]} \BibitemShut
  {NoStop}%
\bibitem [{\citenamefont {Sperhake}\ \emph {et~al.}(2017)\citenamefont
  {Sperhake}, \citenamefont {Moore}, \citenamefont {Rosca}, \citenamefont
  {Agathos}, \citenamefont {Gerosa},\ and\ \citenamefont
  {Ott}}]{Sperhake:2017itk}%
  \BibitemOpen
  \bibfield  {author} {\bibinfo {author} {\bibfnamefont {U.}~\bibnamefont
  {Sperhake}}, \bibinfo {author} {\bibfnamefont {C.~J.}\ \bibnamefont {Moore}},
  \bibinfo {author} {\bibfnamefont {R.}~\bibnamefont {Rosca}}, \bibinfo
  {author} {\bibfnamefont {M.}~\bibnamefont {Agathos}}, \bibinfo {author}
  {\bibfnamefont {D.}~\bibnamefont {Gerosa}}, \ and\ \bibinfo {author}
  {\bibfnamefont {C.~D.}\ \bibnamefont {Ott}},\ }\bibfield  {title} {\enquote
  {\bibinfo {title} {{Long-lived inverse chirp signals from core collapse in
  massive scalar-tensor gravity}},}\ }\href {\doibase
  10.1103/PhysRevLett.119.201103} {\bibfield  {journal} {\bibinfo  {journal}
  {Phys. Rev. Lett.}\ }\textbf {\bibinfo {volume} {119}},\ \bibinfo {pages}
  {201103} (\bibinfo {year} {2017})},\ \Eprint
  {http://arxiv.org/abs/1708.03651} {arXiv:1708.03651 [gr-qc]} \BibitemShut
  {NoStop}%
\bibitem [{\citenamefont {Staykov}\ \emph {et~al.}(2018)\citenamefont
  {Staykov}, \citenamefont {Popchev}, \citenamefont {Doneva},\ and\
  \citenamefont {Yazadjiev}}]{Staykov:2018hhc}%
  \BibitemOpen
  \bibfield  {author} {\bibinfo {author} {\bibfnamefont {K.~V.}\ \bibnamefont
  {Staykov}}, \bibinfo {author} {\bibfnamefont {D.}~\bibnamefont {Popchev}},
  \bibinfo {author} {\bibfnamefont {D.~D.}\ \bibnamefont {Doneva}}, \ and\
  \bibinfo {author} {\bibfnamefont {S.~S.}\ \bibnamefont {Yazadjiev}},\
  }\bibfield  {title} {\enquote {\bibinfo {title} {{Static and slowly rotating
  neutron stars in scalar\textendash{}tensor theory with self-interacting
  massive scalar field}},}\ }\href {\doibase 10.1140/epjc/s10052-018-6064-x}
  {\bibfield  {journal} {\bibinfo  {journal} {Eur. Phys. J. C}\ }\textbf
  {\bibinfo {volume} {78}},\ \bibinfo {pages} {586} (\bibinfo {year} {2018})},\
  \Eprint {http://arxiv.org/abs/1805.07818} {arXiv:1805.07818 [gr-qc]}
  \BibitemShut {NoStop}%
\bibitem [{\citenamefont {Rosca-Mead}\ \emph {et~al.}(2020)\citenamefont
  {Rosca-Mead}, \citenamefont {Moore}, \citenamefont {Sperhake}, \citenamefont
  {Agathos},\ and\ \citenamefont {Gerosa}}]{Rosca-Mead:2020bzt}%
  \BibitemOpen
  \bibfield  {author} {\bibinfo {author} {\bibfnamefont {R.}~\bibnamefont
  {Rosca-Mead}}, \bibinfo {author} {\bibfnamefont {C.~J.}\ \bibnamefont
  {Moore}}, \bibinfo {author} {\bibfnamefont {U.}~\bibnamefont {Sperhake}},
  \bibinfo {author} {\bibfnamefont {M.}~\bibnamefont {Agathos}}, \ and\
  \bibinfo {author} {\bibfnamefont {D.}~\bibnamefont {Gerosa}},\ }\bibfield
  {title} {\enquote {\bibinfo {title} {{Structure of neutron stars in massive
  scalar-tensor gravity}},}\ }\href {\doibase 10.3390/sym12091384} {\bibfield
  {journal} {\bibinfo  {journal} {Symmetry}\ }\textbf {\bibinfo {volume}
  {12}},\ \bibinfo {pages} {1384} (\bibinfo {year} {2020})},\ \Eprint
  {http://arxiv.org/abs/2007.14429} {arXiv:2007.14429 [gr-qc]} \BibitemShut
  {NoStop}%
\bibitem [{\citenamefont {Kuan}\ \emph {et~al.}(2023)\citenamefont {Kuan},
  \citenamefont {Van~Aelst}, \citenamefont {Lam},\ and\ \citenamefont
  {Shibata}}]{Kuan:2023hrh}%
  \BibitemOpen
  \bibfield  {author} {\bibinfo {author} {\bibfnamefont {H.-J.}\ \bibnamefont
  {Kuan}}, \bibinfo {author} {\bibfnamefont {K.}~\bibnamefont {Van~Aelst}},
  \bibinfo {author} {\bibfnamefont {A.~T.-L.}\ \bibnamefont {Lam}}, \ and\
  \bibinfo {author} {\bibfnamefont {M.}~\bibnamefont {Shibata}},\ }\bibfield
  {title} {\enquote {\bibinfo {title} {{Binary neutron star mergers in massive
  scalar-tensor theory: Quasiequilibrium states and dynamical enhancement of
  the scalarization}},}\ }\href {\doibase 10.1103/PhysRevD.108.064057}
  {\bibfield  {journal} {\bibinfo  {journal} {Phys. Rev. D}\ }\textbf {\bibinfo
  {volume} {108}},\ \bibinfo {pages} {064057} (\bibinfo {year} {2023})},\
  \Eprint {http://arxiv.org/abs/2309.01709} {arXiv:2309.01709 [gr-qc]}
  \BibitemShut {NoStop}%
\bibitem [{\citenamefont {Lam}\ \emph {et~al.}(2024)\citenamefont {Lam},
  \citenamefont {Kuan}, \citenamefont {Shibata}, \citenamefont {Van~Aelst},\
  and\ \citenamefont {Kiuchi}}]{Lam:2024wpq}%
  \BibitemOpen
  \bibfield  {author} {\bibinfo {author} {\bibfnamefont {A.~T.-L.}\
  \bibnamefont {Lam}}, \bibinfo {author} {\bibfnamefont {H.-J.}\ \bibnamefont
  {Kuan}}, \bibinfo {author} {\bibfnamefont {M.}~\bibnamefont {Shibata}},
  \bibinfo {author} {\bibfnamefont {K.}~\bibnamefont {Van~Aelst}}, \ and\
  \bibinfo {author} {\bibfnamefont {K.}~\bibnamefont {Kiuchi}},\ }\bibfield
  {title} {\enquote {\bibinfo {title} {{Binary neutron star mergers in massive
  scalar-tensor theory: Properties of post-merger remnants}},}\ }\href@noop {}
  {\  (\bibinfo {year} {2024})},\ \Eprint {http://arxiv.org/abs/2406.05211}
  {arXiv:2406.05211 [gr-qc]} \BibitemShut {NoStop}%
\bibitem [{\citenamefont {Degollado}\ \emph {et~al.}(2020)\citenamefont
  {Degollado}, \citenamefont {Salgado},\ and\ \citenamefont
  {Alcubierre}}]{Degollado:2020lsa}%
  \BibitemOpen
  \bibfield  {author} {\bibinfo {author} {\bibfnamefont {J.~C.}\ \bibnamefont
  {Degollado}}, \bibinfo {author} {\bibfnamefont {M.}~\bibnamefont {Salgado}},
  \ and\ \bibinfo {author} {\bibfnamefont {M.}~\bibnamefont {Alcubierre}},\
  }\bibfield  {title} {\enquote {\bibinfo {title} {{On the formation of
  \textquotedblleft{}supermassive\textquotedblright{} neutron stars and
  dynamical transition to spontaneous scalarization}},}\ }\href {\doibase
  10.1016/j.physletb.2020.135666} {\bibfield  {journal} {\bibinfo  {journal}
  {Phys. Lett. B}\ }\textbf {\bibinfo {volume} {808}},\ \bibinfo {pages}
  {135666} (\bibinfo {year} {2020})},\ \Eprint
  {http://arxiv.org/abs/2008.10683} {arXiv:2008.10683 [gr-qc]} \BibitemShut
  {NoStop}%
\bibitem [{\citenamefont {Mendes}\ and\ \citenamefont
  {Ortiz}(2016)}]{Mendes:2016fby}%
  \BibitemOpen
  \bibfield  {author} {\bibinfo {author} {\bibfnamefont {R.~F.~P.}\
  \bibnamefont {Mendes}}\ and\ \bibinfo {author} {\bibfnamefont
  {N.}~\bibnamefont {Ortiz}},\ }\bibfield  {title} {\enquote {\bibinfo {title}
  {{Highly compact neutron stars in scalar-tensor theories of gravity:
  Spontaneous scalarization versus gravitational collapse}},}\ }\href {\doibase
  10.1103/PhysRevD.93.124035} {\bibfield  {journal} {\bibinfo  {journal} {Phys.
  Rev. D}\ }\textbf {\bibinfo {volume} {93}},\ \bibinfo {pages} {124035}
  (\bibinfo {year} {2016})},\ \Eprint {http://arxiv.org/abs/1604.04175}
  {arXiv:1604.04175 [gr-qc]} \BibitemShut {NoStop}%
\bibitem [{\citenamefont {Ruiz}\ \emph {et~al.}(2012)\citenamefont {Ruiz},
  \citenamefont {Degollado}, \citenamefont {Alcubierre}, \citenamefont
  {N\'u{\~n}ez},\ and\ \citenamefont {Salgado}}]{Ruiz:2012jt}%
  \BibitemOpen
  \bibfield  {author} {\bibinfo {author} {\bibfnamefont {M.}~\bibnamefont
  {Ruiz}}, \bibinfo {author} {\bibfnamefont {J.~C.}\ \bibnamefont {Degollado}},
  \bibinfo {author} {\bibfnamefont {M.}~\bibnamefont {Alcubierre}}, \bibinfo
  {author} {\bibfnamefont {D.}~\bibnamefont {N\'u{\~n}ez}}, \ and\ \bibinfo
  {author} {\bibfnamefont {M.}~\bibnamefont {Salgado}},\ }\bibfield  {title}
  {\enquote {\bibinfo {title} {{Induced scalarization in boson stars and scalar
  gravitational radiation}},}\ }\href {\doibase 10.1103/PhysRevD.86.104044}
  {\bibfield  {journal} {\bibinfo  {journal} {Phys.Rev.}\ }\textbf {\bibinfo
  {volume} {D86}},\ \bibinfo {pages} {104044} (\bibinfo {year} {2012})},\
  \Eprint {http://arxiv.org/abs/1207.6142} {arXiv:1207.6142 [gr-qc]}
  \BibitemShut {NoStop}%
\bibitem [{\citenamefont {Bertotti}\ \emph {et~al.}(2003)\citenamefont
  {Bertotti}, \citenamefont {Iess},\ and\ \citenamefont
  {Tortora}}]{Bertotti:2003rm}%
  \BibitemOpen
  \bibfield  {author} {\bibinfo {author} {\bibfnamefont {B.}~\bibnamefont
  {Bertotti}}, \bibinfo {author} {\bibfnamefont {L.}~\bibnamefont {Iess}}, \
  and\ \bibinfo {author} {\bibfnamefont {P.}~\bibnamefont {Tortora}},\
  }\bibfield  {title} {\enquote {\bibinfo {title} {{A test of general
  relativity using radio links with the Cassini spacecraft}},}\ }\href
  {\doibase 10.1038/nature01997} {\bibfield  {journal} {\bibinfo  {journal}
  {Nature}\ }\textbf {\bibinfo {volume} {425}},\ \bibinfo {pages} {374--376}
  (\bibinfo {year} {2003})}\BibitemShut {NoStop}%
\bibitem [{\citenamefont {Ma}\ \emph {et~al.}(2023)\citenamefont {Ma},
  \citenamefont {Varma}, \citenamefont {Stein}, \citenamefont {Foucart},
  \citenamefont {Duez}, \citenamefont {Kidder}, \citenamefont {Pfeiffer},\ and\
  \citenamefont {Scheel}}]{Ma:2023sok}%
  \BibitemOpen
  \bibfield  {author} {\bibinfo {author} {\bibfnamefont {S.}~\bibnamefont
  {Ma}}, \bibinfo {author} {\bibfnamefont {V.}~\bibnamefont {Varma}}, \bibinfo
  {author} {\bibfnamefont {L.~C.}\ \bibnamefont {Stein}}, \bibinfo {author}
  {\bibfnamefont {F.}~\bibnamefont {Foucart}}, \bibinfo {author} {\bibfnamefont
  {M.~D.}\ \bibnamefont {Duez}}, \bibinfo {author} {\bibfnamefont {L.~E.}\
  \bibnamefont {Kidder}}, \bibinfo {author} {\bibfnamefont {H.~P.}\
  \bibnamefont {Pfeiffer}}, \ and\ \bibinfo {author} {\bibfnamefont {M.~A.}\
  \bibnamefont {Scheel}},\ }\bibfield  {title} {\enquote {\bibinfo {title}
  {{Numerical simulations of black hole-neutron star mergers in scalar-tensor
  gravity}},}\ }\href {\doibase 10.1103/PhysRevD.107.124051} {\bibfield
  {journal} {\bibinfo  {journal} {Phys. Rev. D}\ }\textbf {\bibinfo {volume}
  {107}},\ \bibinfo {pages} {124051} (\bibinfo {year} {2023})},\ \Eprint
  {http://arxiv.org/abs/2304.11836} {arXiv:2304.11836 [gr-qc]} \BibitemShut
  {NoStop}%
\bibitem [{\citenamefont {Mendes}\ \emph {et~al.}(2021)\citenamefont {Mendes},
  \citenamefont {Ortiz},\ and\ \citenamefont {Stergioulas}}]{Mendes:2021fon}%
  \BibitemOpen
  \bibfield  {author} {\bibinfo {author} {\bibfnamefont {R.~F.~P.}\
  \bibnamefont {Mendes}}, \bibinfo {author} {\bibfnamefont {N.}~\bibnamefont
  {Ortiz}}, \ and\ \bibinfo {author} {\bibfnamefont {N.}~\bibnamefont
  {Stergioulas}},\ }\bibfield  {title} {\enquote {\bibinfo {title} {{Nonlinear
  dynamics of oscillating neutron stars in scalar-tensor gravity}},}\ }\href
  {\doibase 10.1103/PhysRevD.104.104036} {\bibfield  {journal} {\bibinfo
  {journal} {Phys. Rev. D}\ }\textbf {\bibinfo {volume} {104}},\ \bibinfo
  {pages} {104036} (\bibinfo {year} {2021})},\ \Eprint
  {http://arxiv.org/abs/2107.07036} {arXiv:2107.07036 [gr-qc]} \BibitemShut
  {NoStop}%
\bibitem [{\citenamefont {Alcubierre}\ \emph {et~al.}(2019)\citenamefont
  {Alcubierre}, \citenamefont {Barranco}, \citenamefont {Bernal}, \citenamefont
  {Degollado}, \citenamefont {Diez-Tejedor}, \citenamefont {Megevand},
  \citenamefont {Núñez},\ and\ \citenamefont {Sarbach}}]{Alcubierre:2019qnh}%
  \BibitemOpen
  \bibfield  {author} {\bibinfo {author} {\bibfnamefont {M.}~\bibnamefont
  {Alcubierre}}, \bibinfo {author} {\bibfnamefont {J.}~\bibnamefont
  {Barranco}}, \bibinfo {author} {\bibfnamefont {A.}~\bibnamefont {Bernal}},
  \bibinfo {author} {\bibfnamefont {J.~C.}\ \bibnamefont {Degollado}}, \bibinfo
  {author} {\bibfnamefont {A.}~\bibnamefont {Diez-Tejedor}}, \bibinfo {author}
  {\bibfnamefont {M.}~\bibnamefont {Megevand}}, \bibinfo {author}
  {\bibfnamefont {D.}~\bibnamefont {Núñez}}, \ and\ \bibinfo {author}
  {\bibfnamefont {O.}~\bibnamefont {Sarbach}},\ }\bibfield  {title} {\enquote
  {\bibinfo {title} {{Dynamical evolutions of $\ell$-boson stars in spherical
  symmetry}},}\ }\href {\doibase 10.1088/1361-6382/ab4726} {\bibfield
  {journal} {\bibinfo  {journal} {Class. Quant. Grav.}\ }\textbf {\bibinfo
  {volume} {36}},\ \bibinfo {pages} {215013} (\bibinfo {year} {2019})},\
  \Eprint {http://arxiv.org/abs/1906.08959} {arXiv:1906.08959 [gr-qc]}
  \BibitemShut {NoStop}%
\bibitem [{\citenamefont {Mendes}\ and\ \citenamefont
  {Ortiz}(2018)}]{Mendes:2018qwo}%
  \BibitemOpen
  \bibfield  {author} {\bibinfo {author} {\bibfnamefont {R.~F.~P.}\
  \bibnamefont {Mendes}}\ and\ \bibinfo {author} {\bibfnamefont
  {N.}~\bibnamefont {Ortiz}},\ }\bibfield  {title} {\enquote {\bibinfo {title}
  {{New class of quasinormal modes of neutron stars in scalar-tensor
  gravity}},}\ }\href {\doibase 10.1103/PhysRevLett.120.201104} {\bibfield
  {journal} {\bibinfo  {journal} {Phys. Rev. Lett.}\ }\textbf {\bibinfo
  {volume} {120}},\ \bibinfo {pages} {201104} (\bibinfo {year} {2018})},\
  \Eprint {http://arxiv.org/abs/1802.07847} {arXiv:1802.07847 [gr-qc]}
  \BibitemShut {NoStop}%
\bibitem [{\citenamefont {Alsing}\ \emph {et~al.}(2012)\citenamefont {Alsing},
  \citenamefont {Berti}, \citenamefont {Will},\ and\ \citenamefont
  {Zaglauer}}]{Alsing2012}%
  \BibitemOpen
  \bibfield  {author} {\bibinfo {author} {\bibfnamefont {J.}~\bibnamefont
  {Alsing}}, \bibinfo {author} {\bibfnamefont {E.}~\bibnamefont {Berti}},
  \bibinfo {author} {\bibfnamefont {C.~M.}\ \bibnamefont {Will}}, \ and\
  \bibinfo {author} {\bibfnamefont {H.}~\bibnamefont {Zaglauer}},\ }\bibfield
  {title} {\enquote {\bibinfo {title} {{Gravitational radiation from compact
  binary systems in the massive {B}rans-{D}icke theory of gravity}},}\ }\href
  {\doibase 10.1103/PhysRevD.85.064041} {\bibfield  {journal} {\bibinfo
  {journal} {Phys. Rev. D}\ }\textbf {\bibinfo {volume} {85}},\ \bibinfo
  {pages} {064041} (\bibinfo {year} {2012})},\ \Eprint
  {http://arxiv.org/abs/1112.4903} {arXiv:1112.4903 [gr-qc]} \BibitemShut
  {NoStop}%
\bibitem [{\citenamefont {Cardoso}\ \emph {et~al.}(2013)\citenamefont
  {Cardoso}, \citenamefont {Carucci}, \citenamefont {Pani},\ and\ \citenamefont
  {Sotiriou}}]{Cardoso:2013fwa}%
  \BibitemOpen
  \bibfield  {author} {\bibinfo {author} {\bibfnamefont {V.}~\bibnamefont
  {Cardoso}}, \bibinfo {author} {\bibfnamefont {I.~P.}\ \bibnamefont
  {Carucci}}, \bibinfo {author} {\bibfnamefont {P.}~\bibnamefont {Pani}}, \
  and\ \bibinfo {author} {\bibfnamefont {T.~P.}\ \bibnamefont {Sotiriou}},\
  }\bibfield  {title} {\enquote {\bibinfo {title} {{Black holes with
  surrounding matter in scalar-tensor theories}},}\ }\href {\doibase
  10.1103/PhysRevLett.111.111101} {\bibfield  {journal} {\bibinfo  {journal}
  {Phys. Rev. Lett.}\ }\textbf {\bibinfo {volume} {111}},\ \bibinfo {pages}
  {111101} (\bibinfo {year} {2013})},\ \Eprint {http://arxiv.org/abs/1308.6587}
  {arXiv:1308.6587 [gr-qc]} \BibitemShut {NoStop}%
\bibitem [{\citenamefont {Brito}\ \emph
  {et~al.}(2015{\natexlab{a}})\citenamefont {Brito}, \citenamefont {Cardoso},\
  and\ \citenamefont {Pani}}]{Brito:2014wla}%
  \BibitemOpen
  \bibfield  {author} {\bibinfo {author} {\bibfnamefont {R.}~\bibnamefont
  {Brito}}, \bibinfo {author} {\bibfnamefont {V.}~\bibnamefont {Cardoso}}, \
  and\ \bibinfo {author} {\bibfnamefont {P.}~\bibnamefont {Pani}},\ }\bibfield
  {title} {\enquote {\bibinfo {title} {{Black holes as particle detectors:
  evolution of superradiant instabilities}},}\ }\href {\doibase
  10.1088/0264-9381/32/13/134001} {\bibfield  {journal} {\bibinfo  {journal}
  {Class. Quant. Grav.}\ }\textbf {\bibinfo {volume} {32}},\ \bibinfo {pages}
  {134001} (\bibinfo {year} {2015}{\natexlab{a}})},\ \Eprint
  {http://arxiv.org/abs/1411.0686} {arXiv:1411.0686 [gr-qc]} \BibitemShut
  {NoStop}%
\bibitem [{\citenamefont {Brito}\ \emph
  {et~al.}(2015{\natexlab{b}})\citenamefont {Brito}, \citenamefont {Cardoso},\
  and\ \citenamefont {Pani}}]{Brito:2015oca}%
  \BibitemOpen
  \bibfield  {author} {\bibinfo {author} {\bibfnamefont {R.}~\bibnamefont
  {Brito}}, \bibinfo {author} {\bibfnamefont {V.}~\bibnamefont {Cardoso}}, \
  and\ \bibinfo {author} {\bibfnamefont {P.}~\bibnamefont {Pani}},\ }\bibfield
  {title} {\enquote {\bibinfo {title} {{Superradiance}: {New Frontiers in Black
  Hole Physics}},}\ }\href {\doibase 10.1007/978-3-319-19000-6} {\bibfield
  {journal} {\bibinfo  {journal} {Lect. Notes Phys.}\ }\textbf {\bibinfo
  {volume} {906}},\ \bibinfo {pages} {pp.1--237} (\bibinfo {year}
  {2015}{\natexlab{b}})},\ \Eprint {http://arxiv.org/abs/1501.06570}
  {arXiv:1501.06570 [gr-qc]} \BibitemShut {NoStop}%
\end{thebibliography}%

\end{document}